\documentclass[aps,preprint,superscriptaddress]{revtex4-2}

\usepackage[normalem]{ulem}
\usepackage{hyperref}
\usepackage{amsmath}
\usepackage{graphicx}
\usepackage{bm}
\usepackage{xcolor}

\newcommand{\stkout}[1]{\ifmmode\text{\sout{\ensuremath{#1}}}\else\sout{#1}\fi}

\preprint{APS/123-QED}
\begin{document}
	
	\noindent
	
	\title[Email of corresponding author: michael.gerard@ipp.mpg.de]{Characterizing Flux-Surface Shapes in Tokamaks and Quasi-Symmetric Stellarators}
	
	\author{M.J.~Gerard \footnote{Corresponding author: michael.gerard@ipp.mpg.de}}
	
	\affiliation{University of Wisconsin-Madison, Wisconsin 53706, USA}
	\affiliation{Max Planck Institute for Plasma Physics, 17491 Greifswald, Germany}
	
	\author{M.J.~Pueschel}
	
	\affiliation{Dutch Institute for Fundamental Research, 5612 AJ Eindhoven, The Netherlands}
	\affiliation{Eindhoven University of Technology, 5600 MB Eindhoven, The Netherlands}
	\affiliation{Department of Physics \& Astronomy, Ruhr-Universit\"at Bochum, 44780 Bochum, Germany}
	
	\author{S.D.~Stewart}
	\author{H.O.M.~Hillebrecht}
	\author{B.~Geiger}
	
	\affiliation{University of Wisconsin-Madison, Wisconsin 53706, USA}
	
	\begin{abstract}
		Modern stellarator designs routinely attain high levels of magnetic-field quasi-symmetry through flux-surface shaping. Here, we examine different methods for characterizing stellarator flux-surface shapes in a manner analogous to flux-surface shaping in tokamaks. The methods considered use a Fourier analysis to define the shaping modes (e.g.~elongation, triangularity, squareness, etc.) of equilibrium cross-sections. Relative to an axisymmetric equilibrium, the additional degree of freedom in a non-axisymmetric equilibrium manifests as a rotation of each shaping mode about the magnetic axis. This analysis is performed on non-axisymmetric configurations with a high degree of quasi-symmetry and equilibria with varying quasi-symmetry quality from the QUASR database.
		
		One method in particular is shown to reduce shape complexity in quasi-symmetric equilibria by defining a set of cross-sections that efficiently fill out an equilibrium volume. This is accomplished by defining a cross-section as the set of points that occupy the shortest distance between the magnetic axis and an equilibrium flux-surface across all quasi-symmetry contours. Using this method, we find empirically an equilibrium geometry can be described with significantly fewer non-negligible shaping modes relative to other shape characterization methods. Moreover, the method reveals that quasi-symmetry quality is strongly correlated with equilibrium shapes that exhibit a highly constrained linear distribution of shaping modes, where an increase in shape complexity is proportional to an increase in shape rotation about the magnetic axis. It is therefore argued this method provides a way to efficiently characterize the shape of quasi-symmetric equilibria in a manner analogous to how equilibrium shapes are described in tokamaks.
	\end{abstract}
	
	\maketitle
	
	\section{Introduction \label{sec:introduction}}
	
	In magnetic confinement fusion, the design and shape of the magnetic field plays a significant role in plasma confinement, influencing a wide range of transport properties that includes magnetohydrodynamic (MHD) stability \cite{Hofmann_1994, Weisen_1997, Weisen2_1997, Scarabosio_2007, Strumberger_2014, hegna_2015, Brunetti_2020, Rodríguez_JPP_2023}, fast-ion confinement \cite{Ghai_2021, Bindel_2023}, neoclassical transport \cite{beidler_demonstration_2021, canik_reduced_2007, canik_experimental_2007}, and turbulence \cite{Angelino_2009, mynick_optimizing_2010, xanthopoulos_controlling_2014, proll_tem_2015, hegna_theory_2018, terry_2018, Fontana_2018, Huang_2019, Laribi_2021, Nakata_2022, Mackenbach_2022, roberg-clark_2022, Merlo_2023, Gerard_2023, Gerard_2024, Stagni_2024, Goodman_2024}. In tokamaks, where the magnetic field is axisymmetric, the relationship between magnetic geometry and plasma dynamics can be investigated using, for instance, the so-called Miller parameters, which provide a geometric description of the axisymmetric poloidal cross-section with parameters like elongation, triangularity, and squareness \cite{Miller_1998}. Importantly, this approach has allowed for the systematic investigation into the effect of shaping on the plasma dynamics in both experiment and simulation \cite{Weisen_1997, Waltz_1999, Reimerdes_2000, Scarabosio_2007, Ball_2017, Fontana_2018, Marinoni2021, Beeke_2021}.
	
	In the (non-axisymmetric) stellarator, equilibrium shapes are typically represented with a Fourier description of their flux surfaces \cite{Nuhrenberg_1986, Nuhrenberg_1988, Garabedian_2008, Dudt_2020, Henneberg_2021}. Such a representation, however, requires spectral condensation or exponential scaling methods to improve convergence \cite{Hirshman_1985, Hirshman_1998, Jang_2026}, and these methods still require $\sim 10^2$ modes. This can make a systematic comparison of the three-dimensional (3D) equilibrium shape between non-axisymmetric configurations challenging. 
	
	From an analytic approach, important insights into the relationship between 3D equilibrium shapes and various figures of merit have been gained by investigating particular asymptotic limits; notably, these include the near-axis-expansion \cite{Mercier_1964, solov_1970, Lortz_1976, Boozer_1991, Landreman_2022, Rodríguez_PPCF_2023} and flux-surface-expansion \cite{hegna_local_2000, Boozer_2002, Skovoroda_2009, Candy_2015, Liu_2026} models. However, neither of these models have provided a geometric description that can fully capture the complexity of stellarator shapes. Alternatively, some bespoke descriptions of plasma shaping have been shown to predict changes in microinstability dynamics in the Helically Symmetric eXperiment (HSX) stellarator \cite{Gerard_2023, Gerard_2024}. These methods, however, also fail to provide a generalizable method for characterizing stellarator shapes.
	
	Here, three different methods for characterizing a global ideal MHD equilibrium shape are compared. It is found, empirically, that one of these methods results in $\sim 10$ non-negligable shaping modes in non-axisymmetric equilibria. Notably, this reduction is found across equilibria with a wide range of quasi-symmetry quality, though the reduction tends to increase as quasi-symmetry improves. Additionally, the $\sim 10$ shaping modes are found to be more tightly constrained with improved quasi-symmetry. 
	
	The shaping methods considered are based on one of two different definitions for a poloidally closed cross-section. The first set of cross-sections is defined to be locally perpendicular to the magnetic axis and is referred to as planar cross-sections. The second set is comprised of the points occupying the shortest distance between the magnetic axis and the plasma edge while adhering to contours of quasi-symmetry. This set is referred to as the torsioned cross-sections, and they are argued to efficiently fill out the volume of a quasi-symmetric equilibrium. Notably, both cross-section definitions result in identical poloidal cross-sections when applied to an axisymmetric system. The three methods for characterizing an equilibrium shape are defined using a two-dimensional (2D) Fourier transform of these cross-sections in two different sets of coordinates. Two of these methods are based on the planar cross-sections and one is based on the torsioned cross-sections. A discussion on axisymmetry and quasi-symmetry, the cross-section definitions, and how they are used to characterize an equilibrium shape are presented in Sec.~\ref{sec:defining_cross_sections}.
	
	In Sec.~\ref{sec:non-axisymmetry}, three shape characterization methods are compared in the precise quasi-axisymmetric (QA) and precise quasi-helically symmetric (QH) configurations of Ref.~\cite{landreman_magnetic_2022}. It is shown that the torsioned cross-sections result in a significant reduction in the number of non-negligible spectral modes relative to the planar cross-sections. It is argued that the reduction in spectral content is a consequence of the efficiency with which the torsioned cross-sections fill out the volume of a quasi-symmetric equilibrium. It is also shown that the precise QA and QH configurations exhibit a highly constrained set of shaping modes in their shape spectrum. This trend is most clearly visible in the analysis performed with the torosioned cross-sections.
	
	These trends are similarly observed in Sec.~\ref{sec:QUASR}, where all three shape characterization methods are compared throughout a large subset of the quasi-symmetric Stellarator Repository (QUASR), which hosts over $3\times 10^5$ QA and QH stellarator equilibria \cite{Giuliani_2024, giuliani_2024_13717741, Giuliani_2025}.	It is found that the reduction in spectral content associated with the torsioned cross-sections persists even when the quasi-symmetry is significantly degraded. Moreover, the degradation is observed to correlate strongly with how the spectral shaping modes are distributed. It is found that equilibria with good quasi-symmetry produce shape spectra with a linear distribution of shaping modes while equilibria with poor quasi-symmetry have a more isotropic distribution. Based on these observations, it is inferred that an increase in shape complexity (i.e., higher-order shaping modes) requires a proportional increase in the rate at which those shapes rotate about the magnetic axis, and that deviations from such a distribution result in a significant degradation in quasi-symmetry. 
		
	Based on these observations, it is concluded in Sec.~\ref{sec:conclusion} that the torsioned cross-sections provide a method for efficiently characterizing a quasi-symmetric equilibrium shape by reducing the spectral content and constraining how the remaining content is distributed. This is argued to be a consequence of the efficiency with which the torsioned cross-sections fill out the volume of a quasi-symmetric equilibrium. It is therefore argued that, similar to how the corresponding poloidal cross-sections in an axisymmetric system are used to systematically investigate the relationship between flux-surface shaping and transport, so too can the torsioned cross-sections be used in quasi-symmetric equilibria. 
	
	\section{Cross-sections in Non-axisymmetric Equilibria \label{sec:defining_cross_sections}}
	
	When considering how one might define a generalized shape description, it is first helpful to consider that axisymmetry is defined as a spatial symmetry along the magnetic axis. This means that all poloidal cross-sections are identical, where a poloidal cross-section is defined as a poloidally closed flux-surface path that follows a contour in the toroidal angle $\varphi$, where $\varphi$ is equivalent to the azimuthal angle in a cylindrical coordinate system. Therefore, in an axisymmetric equilibrium, if one simultaneously knows a single poloidal cross-section, the magnetic axis, and the position of a poloidal cross-section relative to the magnetic axis, then one has a minimal description of the entire equilibrium shape. Such a minimal description then facilitates more systematic investigations into the relationship between flux-surface shaping and transport.
	
	In non-axisymmetric equilibria, generally no such spatial symmetry exists. However, there can be a quasi-symmetry in which there exists an invariant direction in the magnetic field strength $B=|\mathbf{B}|$, provided $B$ is represented in an appropriate coordinate system (e.g., Boozer coordinates) \cite{Boozer_1983, Nuhrenberg_1988, helander_theory_2014}. The torsioned cross-sections considered throughout this manuscript are constructed from the quasi-symmetry contours of such an equilibrium. 
	
	In practice, perfect quasi-symmetry may not exist. However, in modern stellarator design, flux-surface shapes are optimized to produce a high degree of accuracy in approximating a quasi-symmetric magnetic field \cite{Bader_2019, Henneberg_2019, landreman_magnetic_2022, Nies_2022, Giuliani_JPP_2022, Giuliani_JCP_2022, Wechsung_2022, Dudt_2023, Buller_2025}. This highlights an important distinction between the quasi-symmetry contours and the actual contours of the magnetic field strength $B$. Using the poloidal $\theta$ and toroidal $\phi$ Boozer coordinates \cite{boozer_guiding_1980}, a set of quasi-symmetry coordinates
	\begin{align}
		\eta &= M\theta - N\phi, \label{eq:eta_def} \\
		\xi &= \begin{cases}
			\phi & \mathrm{for} \ \mathrm{QA} \ \left(M\neq 0 \ \mathrm{and} \ N=0\right) \\
			\theta/M + \phi/N & \mathrm{for} \ \mathrm{QH} \ \left(M\neq 0 \ \mathrm{and} \ N\neq 0\right)
		\end{cases} \label{eq:xi_def}
	\end{align}
	is introduced. Here, $M$ and $N$ are the poloidal and toroidal mode numbers of the dominant quasi-symmetry mode, respectively. This means the $\eta$ contours define the quasi-symmetry contours and the $\xi$ contours are orthogonal to the $\eta$ contours when represented in Boozer coordinates. Therefore, a quasi-symmetric equilibrium is one in which the $B$ contours are approximately aligned with the $\eta$ contours. Note that quasi-poloidal symmetry (i.e., $M=0$ and $N\neq0$) is neglected, since no QUASR-like database exists for quasi-poloidally symmetric equilibria.
	
	\begin{figure*}
		\centering
		\includegraphics[width=0.98\textwidth, keepaspectratio]{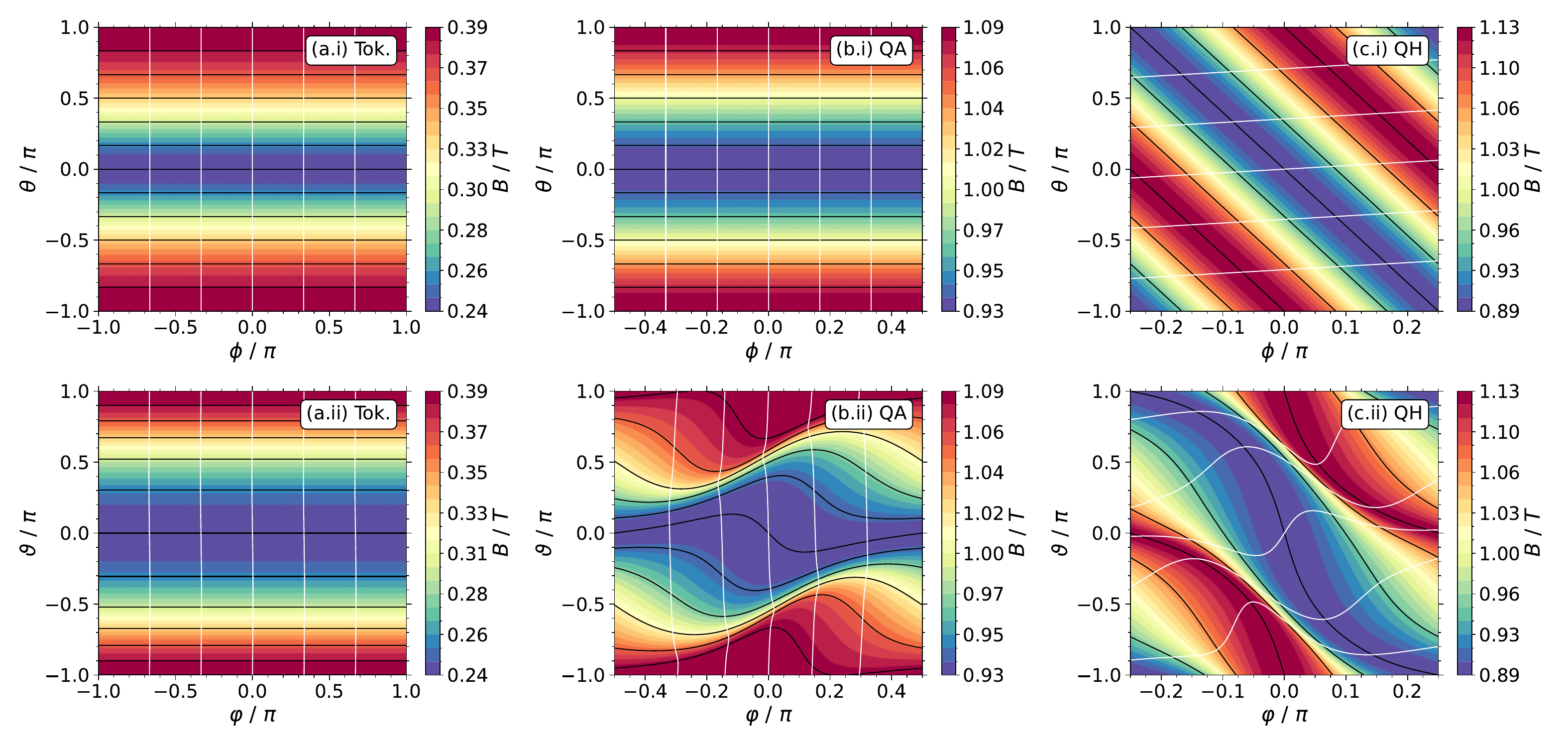}
		\caption{The magnetic field strength is plotted on the $\psi/\psi_\mathrm{edge}=0.5$ flux surface as a function of Boozer angles $\phi$ and $\theta$ in the top row and the and toroidal $\varphi$ and poloidal $\vartheta$ angles in the bottom row. Columns, from left to right, correspond to a circular tokamak, a precise QA, and a precise QH configuration. Included in each panel are the $\eta$ (black) and $\xi$ (white) contours. This shows that while the quasi-symmetry contours are uniformly distributed and orthogonal in Boozer coordinates, neither is generally true in toroidal and poloidal coordinates. \label{fig:sym_coords}}
	\end{figure*}
	
	The $\eta$ and $\xi$ coordinates are shown in Fig.~\ref{fig:sym_coords} for a tokamak with a circular cross-section and the precise QA and QH configurations. In each panel, the magnetic field strength on the $\psi/\psi_\mathrm{edge}=0.5$ flux surface is shown, where $\psi$ is the toroidal magnetic flux used to label each flux surface and $\psi_\mathrm{edge}$ is $\psi$ at the plasma boundary. Each flux surface is plotted over a single field period, and columns, respectively from left to right, correspond to the tokamak, the precise QA, and the precise QH configuration. The top row shows $B$ as a function of $\theta$ and $\phi$, with the $\eta$ and $\xi$ contours shown as black and white lines, respectively. 
	
	Across these panels, the alignment between the $\eta$ contours and the $B$ contours, as well as the orthogonality between the $\xi$ and $\eta$ coordinates, can be readily observed. Note that in the precise QH configuration shown in panel (c.i), the angle between the $\eta$ and $\xi$ contours appears larger than 90 degrees because of the aspect ratio between the poloidal and toroidal domains. This orthogonality, however, is only observed when the $\eta$ and $\xi$ contours are represented in Boozer coordinates; in physical space, the coordinates are not orthogonal. This is demonstrated in the bottom row of Fig.~\ref{fig:sym_coords}, where the same data is shown with respect to the toroidal $\varphi$ and poloidal $\vartheta = \arctan((Z-Z_\mathrm{ma})/(R-R_\mathrm{ma}))$ angles, where $R$ and $Z$ are the radial and vertical coordinates in a cylindrical coordinate system, and the $\mathrm{ma}$ subscript denotes the corresponding variable on the magnetic axis. Notably, curves that are orthogonal when represented with respect to the coordinates $\vartheta$ and $\varphi$ are orthogonal in physical space. Therefore, in this representation, it can be observed that the $\eta$ and $\xi$ contours are orthogonal in the axisymmetric tokamak of panel (a.ii), but not in the non-axisymmetric QA and QH equilibria of panels (b.ii) and (c.ii), respectively. It should also be noted that the $\eta$ contours tend to be more tightly packed in regions where the flux-surface gradient of $B$ is large, which can be observed in panels (b.ii) and (c.ii) around $\varphi=0$ and $\vartheta=\pm \pi/2$. 
	
	When considering how best to define a flux-surface cross-section that will be used to analyze an equilibrium shape, it is assumed that the physical deformation of the $B$ contours contains important geometric information regarding an equilibrium surface. Therefore, the alignment between the $\eta$ and $B$ contours in a quasi-symmetric equilibrium makes $\eta$ a suitable coordinate with which to parameterize a position along a cross-section path.
	%As will be shown in Sec.~\ref{sec:non-axisymmetry}, this choice of cross-section parametrization will result in a significant reduction in non-negligible shaping modes relative to another poloidal parametrization, thereby supporting this hypothesis.
	
	\begin{figure*}
		\includegraphics[width=0.98
		\textwidth, keepaspectratio]{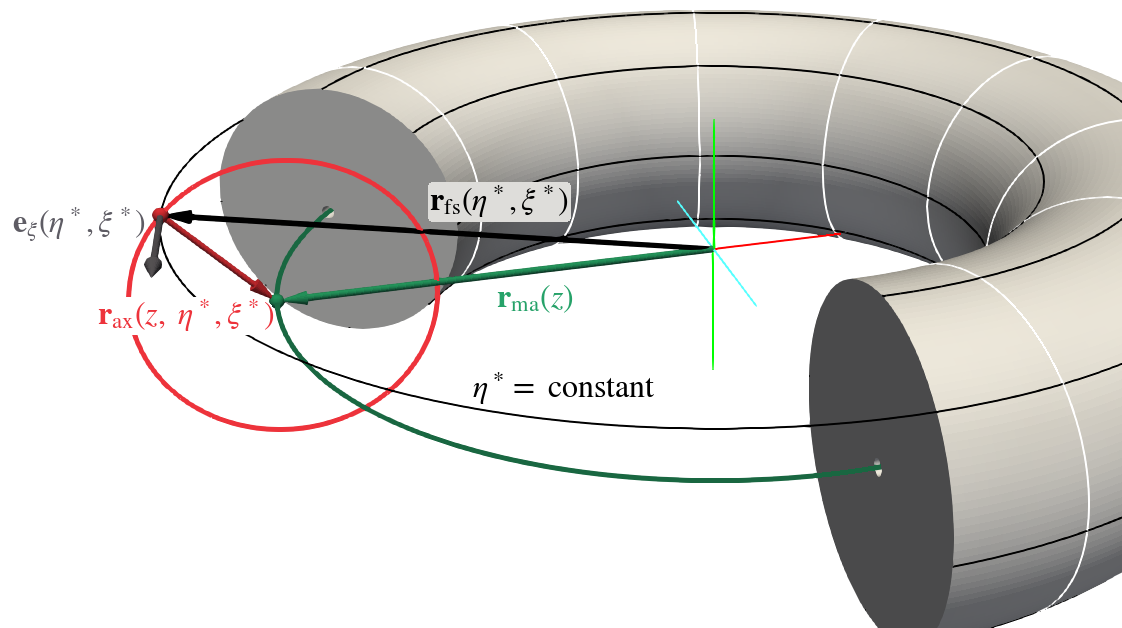}
		\caption{A demonstration of the coordinate system described in the text is shown for a circular-cross-section tokamak. The black and white curves show the $\eta$ and $\xi$ contours, respectively, and the green curve is the magnetic axis. The $\mathbf{e}_{\xi}\cdot \mathbf{r}_\mathrm{ax}=0$ contour centered about the point $\mathbf{r}_\mathrm{ma}(z)$ on the magnetic axis is marked in red. The red point along this contour is the $\eta=\eta^*$ location of the corresponding poloidal cross-section. \label{fig:coord_demo}}
	\end{figure*}
	
	Here, two methods for defining an equilibrium cross-section at a point along the magnetic axis, labeled by $z$, are defined. Both methods will enable the cross-section to close on itself in the poloidal direction after a $2\pi$ change in $\eta$. To allow for other possible parametrizations, the poloidal cross-section coordinate will, for now, be labeled $\lambda$. The vector that extends from $z$ on the axis to trace out each cross-section path will be labeled $\boldsymbol{\rho}(z, \lambda)$. Moreover, if either method is applied to an axisymmetric system, it will result in a axisymmetric poloidal cross-section. 
	
	The first method can be applied to any ideal MHD equilibrium, and is defined by the intersection of a flux surface with a plane perpendicular to the magnetic axis at $z$. These cross-sections will therefore be referred to as planar cross-sections. While simple to define, this method does not explicitly invoke any notion of quasi-symmetry, aside from the fact that one is free to choose $\lambda=\eta$. It will be shown in Secs.~\ref{sec:non-axisymmetry} and \ref{sec:QUASR}, however, that this choice in $\lambda$ results in a significant reduction in non-negligible spectral shaping modes relative to the choice of a geometric poloidal angle.
	
	The second method defines $\boldsymbol{\rho}$ as 
	\begin{equation}
	\begin{split}
			\boldsymbol{\rho}(z, \eta) &= \mathbf{r}_\mathrm{fs}(\eta, \xi^*) - \mathbf{r}_\mathrm{ma}(z), \\
		\xi^*(z, \eta) &= \mathrm{argmin}_{\xi} \left[ r_\mathrm{ax}(z, \eta, \xi) \right], 
	\end{split} \label{eq:cross_section}
	\end{equation}
	where $\mathbf{r}_\mathrm{ma}(z)$ is the vector that points from the origin to the magnetic axis, $\mathbf{r}_\mathrm{fs}(\eta, \xi)$ is the vector that points from the origin to a point on the flux surface, $r_\mathrm{ax}(z, \eta,\xi) = \| \mathbf{r}_\mathrm{ax}(z, \eta, \xi) \|$, with  $\mathbf{r}_\mathrm{ax} = \mathbf{r}_\mathrm{ma}(z) - \mathbf{r}_\mathrm{fs}(\eta, \xi)$ and $\|\cdot\|$ denoting a Euclidean 2-norm, and the $\mathrm{argmin}_{\xi}$ function identifies the value of $\xi$ that minimizes the function argument at fixed $\eta$ and $z$. From this definition, $\boldsymbol{\rho}$ can be understood to be comprised of the set of points across all $\eta$ contours that minimize the distance between each $\eta$ contour and the point $z$ on the magnetic axis. 
	
	To help visualize this coordinate system, Fig.~\ref{fig:coord_demo} shows a circular tokamak with the $\eta$ and $\xi$ contours wrapping around the torus toroidally and poloidally, respectively. An exemplary $\eta$ contour, labeled $\eta^*$, is shown within a toroidal gap, over which the equilibrium volume is cut away to reveal the magnetic axis, which is colored green. The black $\mathbf{r}_\mathrm{fs}$ vector is shown at the particular values of $\eta^*$ and $\xi^*$ at which $r_\mathrm{ax}$ is minimized, with $\mathbf{r}_\mathrm{ax}$ shown in red. Note that at this point, the covariant basis vector $\mathbf{e}_{\xi} = \partial \mathbf{r} /\partial \xi$, which is defined while keeping $\eta$ and $\psi$ fixed, is also shown in dark gray, and, by definition, is tangent to the $\eta^*$ contour. Importantly, at this point of minimal $r_\mathrm{ax}$, the vectors $\mathbf{e}_{\xi}$ and $\mathbf{r}_\mathrm{ax}$ are orthogonal.  
	
	This orthogonality is a direct consequence of the extremization of $r_\mathrm{ax}$. This can be understood by taking the derivative of $r_\mathrm{ax}$ with respect to $\xi$, holding all other variables fixed. Then, setting that derivative to zero, one finds that
	\begin{equation}
		\mathbf{e}_{\xi} \cdot \mathbf{r}_\mathrm{ax} = 0. \label{eq:min_cond_one}
	\end{equation}
	This means that a minimum in $r_\mathrm{ax}$ can only occur when a vector, tangent to a particular $\eta$ contour, is orthogonal to the vector that points from that point on a flux surface to $\mathbf{r}_\mathrm{ma}(z)$. Immediately, one can infer that in an axisymmetric system, where $\xi = \varphi$ and $\mathbf{e}_{\xi}=\mathbf{e}_{\varphi}$, orthogonality between $\mathbf{r}_\mathrm{ax}$ and $\mathbf{e}_{\xi}$ means that $\mathbf{r}_\mathrm{ax}$ must lie within the plane perpendicular to the toroidal direction (i.e.~a polodial cross-section), and this is indeed observed in Fig.~\ref{fig:coord_demo}.
	
	Equation (\ref{eq:min_cond_one}), however, is only sufficient to prove that $r_\mathrm{ax}$ is at a local extremum. A local minimum also requires that $\partial^2r_\mathrm{ax}/\partial\xi^2 > 0$. Therefore, taking the second derivative of $r_\mathrm{ax}$ with respect to $\xi$, and asserting Eq.~(\ref{eq:min_cond_one}), one finds
	\begin{equation}
		\frac{\partial^2r_\mathrm{ax}}{\partial \xi^2} = \frac{1}{r_\mathrm{ax}} \left[ g_{\xi\xi} - \frac{\partial \mathbf{e}_{\xi}}{\partial \xi} \cdot \left( \mathbf{r}_\mathrm{ma} - \mathbf{r}_\mathrm{fs} \right) \right], \label{eq:minimum_intermediate}
	\end{equation}
	where $g_{\xi\xi} = \mathbf{e}_{\xi}\cdot\mathbf{e}_{\xi}$. To then derive an expression for $\partial\mathbf{e}_{\xi}/\partial\xi$, consider that in a Frenet-Serret frame \cite{Dhaeseleer}
	\begin{equation}
		\frac{\partial^2\mathbf{r}_\mathrm{fs}}{\partial s^2} = \kappa \hat{\mathbf{N}} \label{eq:frenet-serret},
	\end{equation}
	where $s$ is the arc-length along the $\eta$ contour traced out by $\mathbf{r}_\mathrm{fs}$, $\kappa$ is the local curvature along that contour, and $\hat{\mathbf{N}}$ is a unit vector orthonormal to $\mathbf{e}_{\xi}$. Expressed in terms of $\xi$, Eq.~(\ref{eq:frenet-serret}) becomes
	\begin{equation}
		\frac{\partial^2\xi}{\partial s^2} \mathbf{e}_{\xi} + \frac{1}{g_{\xi\xi}} \frac{\partial\mathbf{e}_{\xi}}{\partial\xi} = \kappa \hat{\mathbf{N}}, \label{eq:frenet-serret_xi}
	\end{equation}
	using the relation $\partial\xi/\partial s = 1/\sqrt{g_{\xi\xi}}$. Then, rearranging Eq.~(\ref{eq:frenet-serret_xi}) to isolate $\partial\mathbf{e}_{\xi}/\partial\xi$, and substituting into Eq.~(\ref{eq:minimum_intermediate}), one finds
	\begin{equation}
		\frac{\partial^2r_\mathrm{ax}}{\partial\xi^2} = \frac{g_{\xi\xi}}{r_\mathrm{ax}} \left( 1 - \kappa r_\mathrm{ax} \hat{\mathbf{N}} \cdot \hat{\mathbf{r}}_\mathrm{ax} \right), \label{eq:minimum_intermediate_two}
	\end{equation}
	with $\hat{\mathbf{r}}_\mathrm{ax} = \mathbf{r}_\mathrm{ax}/r_\mathrm{ax}$ and $\mathbf{e}_{\xi}\cdot\mathbf{r}_\mathrm{ax} = 0$, where the latter is required to be consistent with Eq.~(\ref{eq:min_cond_one}). In Eq.~(\ref{eq:minimum_intermediate_two}), $g_{\xi\xi}$ and $r_\mathrm{ax}$ are positive definite, $\kappa$ is positive semi-definite, and $\hat{\mathbf{N}}\cdot\hat{\mathbf{r}}_\mathrm{ax} \in [-1,\,1]$. Therefore, independent of the value of the scalar product, $\partial^2r_\mathrm{ax}/\partial\xi^2 > 0$ if
	\begin{equation}
		\kappa < \frac{1}{r_\mathrm{ax}}. \label{eq:min_cond_two}
	\end{equation}
	This means that, in an axisymmetric system, if a poloidal cross-section satisfies Eq.~(\ref{eq:min_cond_two}), then that cross-section is defined as the set of points that locally minimize the distance between each $\eta$ contour and the corresponding point on the magnetic axis. 
	
	To determine whether a poloidal cross-section is likely to satisfy Eq.~(\ref{eq:min_cond_two}), consider that the $\eta$ contours wrap toroidally around a flux surface. It is therefore argued that $\kappa \approx 1/R_0$, where $R_0$ is the major radius. If one then restricts their analysis to only the set of points satisfying Eq.~(\ref{eq:min_cond_one}) that are in the vicinity of the specified axial position $z$, then $r_\mathrm{ax} \approx a$, where $a$ is the minor radius. From these scaling relations, Eq.~(\ref{eq:min_cond_two}) can be approximated as $a/R_0 < 1$, which is always observed. Therefore, the set of points satisfying Eq.~(\ref{eq:min_cond_one}) can be reasonably assumed to be a local minimum, provided they are points within the toroidal vicinity of $z$.
	
	In a non-axisymmetric equilibrium, if there exists a set of points that form a smooth and poloidally closed curve that satisfies Eqs.~(\ref{eq:min_cond_one}) and (\ref{eq:min_cond_two}), then those points would describe a one-dimensional (1D) cross-section over which the distance between $z$ and each $\eta$ contour is minimized. In this way, such a cross-section can be considered analogous to the poloidal cross-section in an axisymmetric system. If one then imagines forming an embedded 2D cross-section by connecting the 1D cross-sections across adjacent flux surfaces---connecting them by their shared $\eta$ values---then, by construction, this cross-section would be composed of the shortest paths between the magnetic axis and the plasma boundary while adhering to surfaces of quasi-symmetry. Therefore, it is argued that these hypothetical 2D cross-sections would provide an efficient way to fill out the volume of a quasi-symmetric equilibrium.
	
	At this point, it is important to stress that the existence of a non-axisymmetric cross-section satisfying Eqs.~(\ref{eq:min_cond_one}) and (\ref{eq:min_cond_two}), that is smooth, and poloidally closed, has not been proven. It is reported, however, that such cross-sections are found to exist everywhere along the magnetic axis of the $217,082$ non-axisymmetric equilibria investigated from the QUASR database. Furthermore, the axisymmetric scaling argument for satisfying Eq.~(\ref{eq:min_cond_two}) by limiting one's analysis to the toroidal vicinity of $z$ is found to be sufficient for these non-axisymmetric equilibria. This is somewhat surprising considering the set includes $2,376$ QH equilibria with $|N/M|=8$, where one might expect $\kappa$ to approach values of $1/a$. But in all such cases, only one poloidally closed and smooth contour of $\mathbf{e}_{\xi} \cdot \mathbf{r}_\mathrm{ax} = 0$ is found in the toroidal vicinity of $z$.
	
	\begin{figure*}
		\centering
		\includegraphics[width=0.98\textwidth, keepaspectratio]{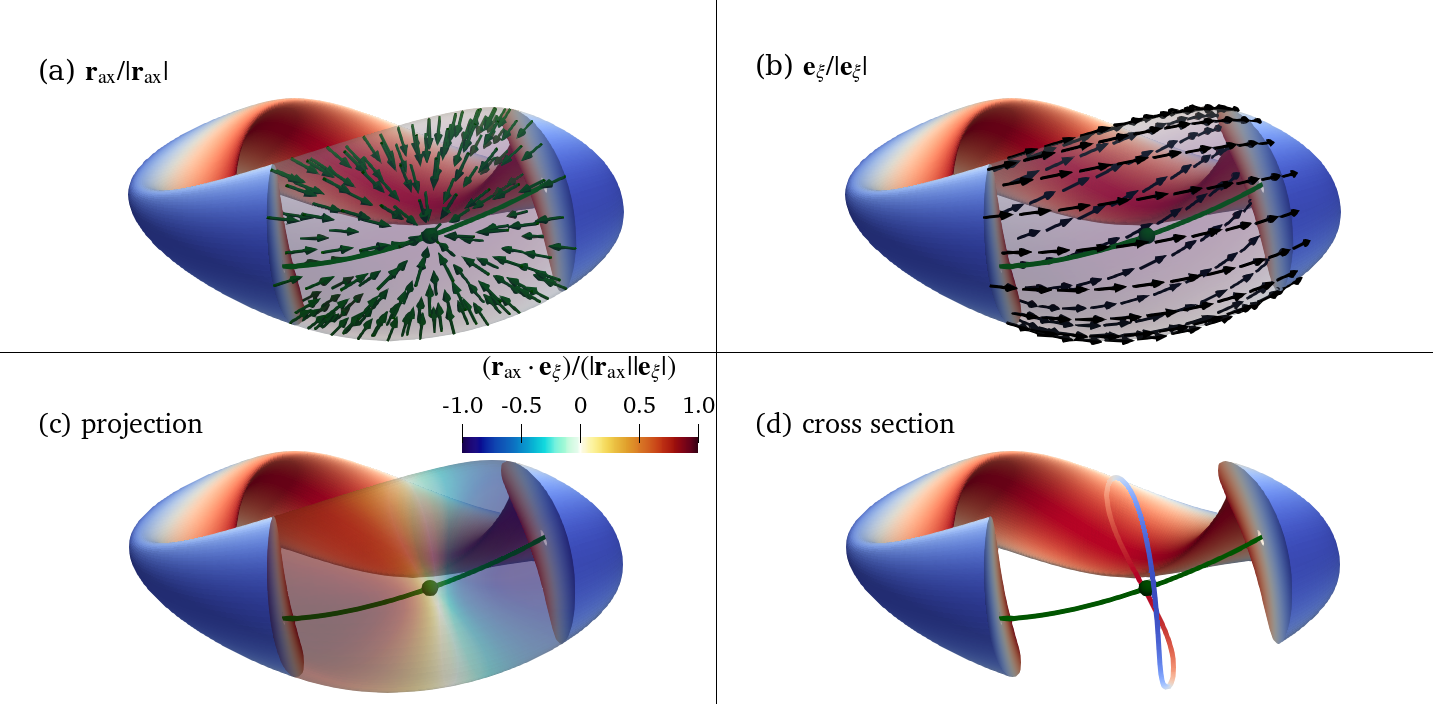}
		\caption{The algorithm for identifying a torsioned cross-section is shown graphically in the precise QA configuration. Panel (a) shows the $\mathbf{r}_\mathrm{ax}$ vector field over a portion of a flux surface, panel (b) shows the covariant $\xi$ basis vector field $\mathbf{e}_{\xi}$, panel (c) shows the projection of these two fields, and panel (d) shows the cross-section described by the roots of the projection. In each panel, the magnetic axis is shown as a green curve, while the point along the magnetic axis where the cross-section is defined is indicated with a green point. \label{fig:QA_cross_section_demo}}
	\end{figure*}
	
	\begin{figure*}
		\centering
		\includegraphics[width=0.98\textwidth, keepaspectratio]{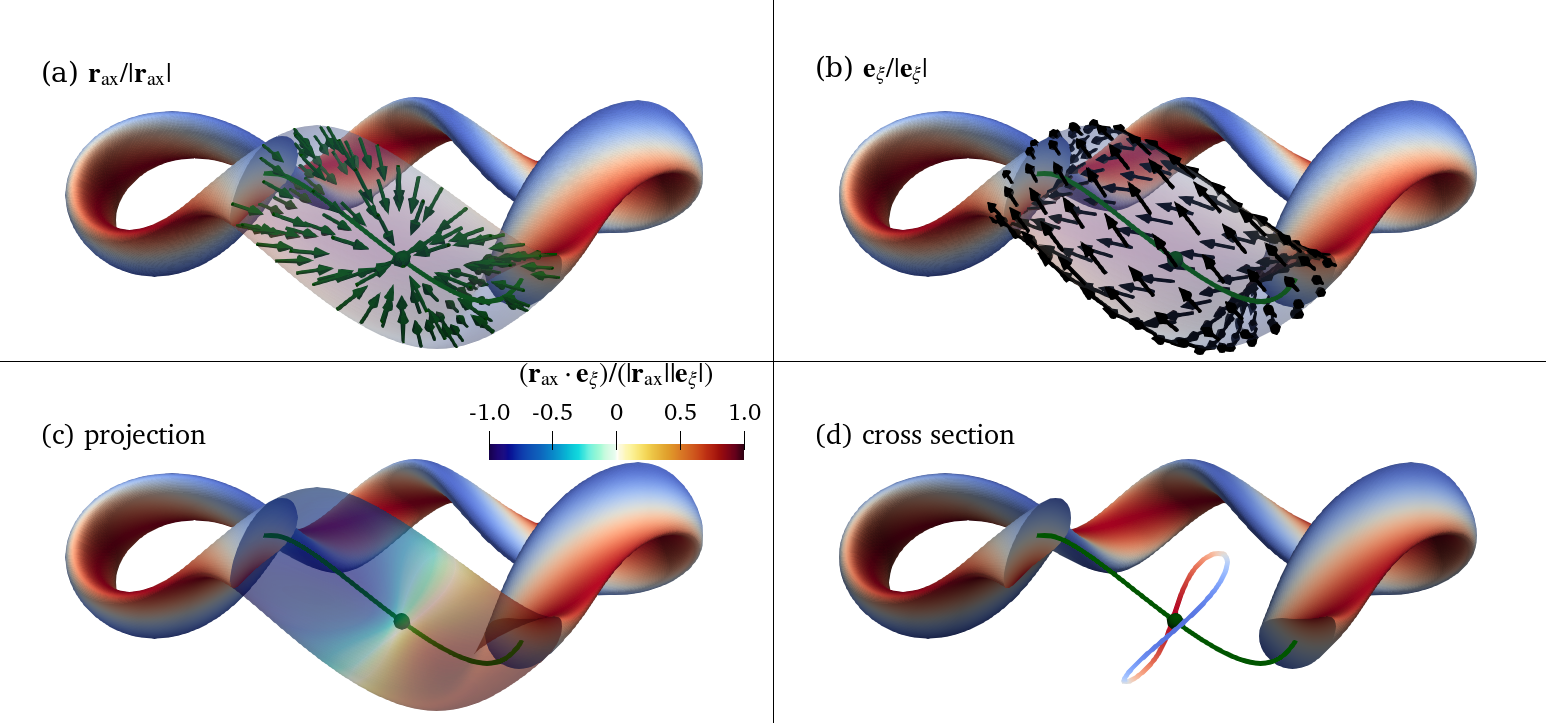}
		\caption{The algorithm for identifying a torsioned cross-section is shown graphically in the precise QH configuration. Panel (a) shows the $\mathbf{r}_\mathrm{ax}$ vector field over a portion of a flux surface, panel (b) shows the covariant $\xi$ basis vector field $\mathbf{e}_{\xi}$, panel (c) shows the projection of these two fields, and panel (d) shows the cross-section described by the roots of the projection. In each panel, the magnetic axis is shown as a green curve, while the point along the magnetic axis where the cross-section is defined is indicated with a green point. \label{fig:QH_cross_section_demo}}
	\end{figure*}
	
	For the cross-sections considered in this paper, the toroidal vicinity is taken to mean a set of points along a flux surface within a half-field-period on either side of some position $z$. The related cross-sections are then computed in three stages. In the first stage, the $\mathbf{r}_\mathrm{ax}$ and $\mathbf{e}_{\xi}$ vector fields are computed on a flux surface over the toroidal vicinity of $z$. Then, the scalar projection between $\mathbf{r}_\mathrm{ax}$ and $\mathbf{e}_{\xi}$ is computed over that same domain. Finally, the $z$ cross-section is identified using a root-finding algorithm to trace out a poloidally closed and smooth path over which $\mathbf{r}_\mathrm{ax}\cdot\mathbf{e}_{\xi}=0$. While this method of defining a cross-section can result in a planar cross-section, it does not do so generally. Therefore, these cross-sections will be referred to as torsioned cross-sections.
	
	Two example calculations are shown in Figs.~\ref{fig:QA_cross_section_demo} and \ref{fig:QH_cross_section_demo}, which correspond to the precise QA and the precise QH configurations, respectively. In each figure, the magnetic axis is represented as a green curve with the specified $z$ position shown with a green point. The $\mathbf{r}_\mathrm{ax}$ vector field is shown in panel (a), the $\mathbf{e}_{\xi}$ vector field is shown in panel (b), the scalar projection between the two is shown in panel (c), and the resulting cross-section is shown in panel (d).

	\subsection{Defining equilibrium shapes via modal analysis}

	In this section, a distinction is made between the cross-section shape $\rho(z,\lambda) = \|\boldsymbol{\rho}\|$ and the cross-section path $\hat{\boldsymbol{\rho}}(z, \lambda) = \boldsymbol{\rho} / \rho$. This choice is made to be consistent with how shaping is defined in an axisymmetric equilibrium, where shaping (i.e., elongation, triangularity, squareness, etc.) is defined exclusively by $\rho$. To characterize the equilibrium shapes, a discrete Fourier transform is performed in $z$ and $\lambda$.

	This transformation is expressed as
	\begin{equation}
		\rho(z, \lambda) = \sum_{k=-\infty}^{\infty} \sum_{\ell=0}^{\infty} \hat{\rho}(k, \ell) \mathrm{e}^{-i\left(k z + \ell\lambda \right)}, \label{eq:fourier_shape}
	\end{equation}
	making the shape spectrum $\hat{\rho}$ a complex-valued function of $k$ and $\ell$. In order for $\rho$ to be a real-valued quantity, $\hat{\rho}(k, \ell) = \hat{\rho}^*(-k, -\ell)$. Therefore, only the $\ell \geq 0$ modes need to be included in Eq.~(\ref{eq:fourier_shape}). 

\begin{figure}
		\centering
		\includegraphics[width=0.48\textwidth, keepaspectratio]{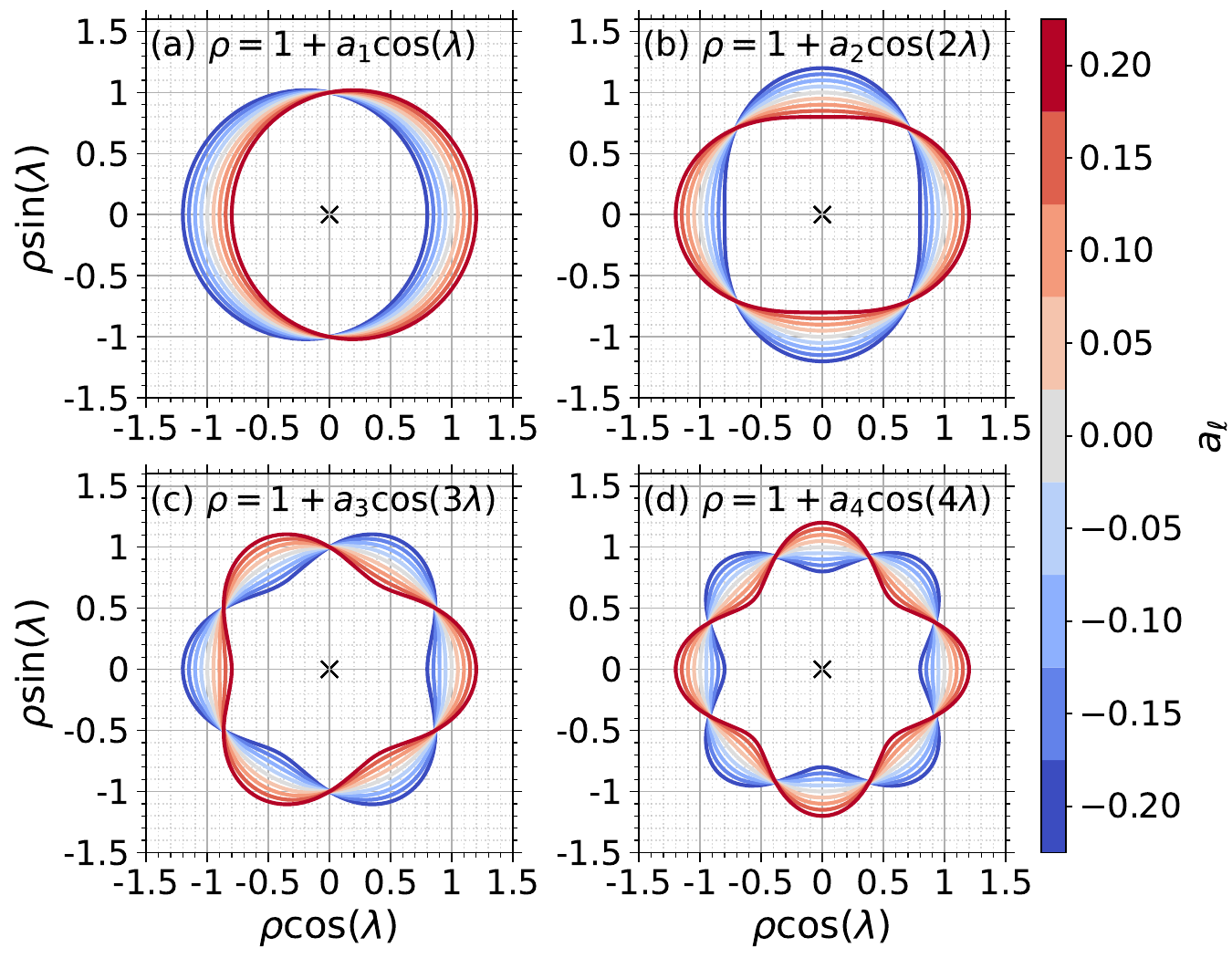}
		\caption{The $\ell=1,\, 2,\ 3,$ and $4$ shaping modes are shown, with mode amplitudes $a_{\ell}$ depicted on the color scale. This provides a pictorial interpretation of how the shaping modes relate to familiar concepts like the Shafranov shift and shaping parameters like elongation, triangularity, and squareness. \label{fig:mode_demo}}
	\end{figure}

	To gain additional insight for interpreting a shape spectrum $\hat{\rho}$, Fig.~\ref{fig:mode_demo} shows a set of hypothetical cross-sections at some $z$ for $\ell=1$, $2$, $3$, and $4$ in panels (a), (b), (c), and (d), respectively. For simplicity, only the cosine terms are included in the demonstration. The color scale shows the shape amplitude $a_{\ell}$, with positive (negative) values of $a_{\ell}$ shaded in red (blue). From panel (a), the $\ell=1$ mode can be related to a Shafranov-like shift, though its more precise interpretation is as a measure of the translation of the geometric center of a cross-section along the $\lambda =0$ axis. This differentiates the present Fourier representation from that of previous publications, where the $\ell=1$ Fourier mode describes the helicity of the magnetic axis \cite{Nuhrenberg_1986, Nuhrenberg_1988, Garabedian_2008}. For $\ell \geq 2$, however, the shaping modes are analogous to the definitions in those citations. This is observed in panels (b), (c), and (d), where the $\ell=2$, $3$, and $4$ modes can be observed to describe an elongation, triangularity, and squareness, respectively. Note that the inclusion of finite-amplitude sine terms would result in a rotation of the corresponding $\ell$ shape relative to the $\lambda=0$ axis. 
	
	In an axisymmetric equilibrium, the shape spectrum $\hat{\rho}$ is populated exclusively by shaping modes with $k=0$, meaning the corresponding shapes undergo no axial rotation as one traverses the magnetic axis. Alternatively, in a non-axisymmetric equilibrium, the $\ell$ shaping modes are free to rotate about the magnetic axis. These rotations, however, must close on themselves over each magnetic field period. Note that in the proceeding sections, $z$ is defined as either $\varphi_\mathrm{ma}$ or $\phi_\mathrm{ma}$, where the subscript denotes the corresponding quantity along the magnetic axis. In either case, this means that $\rho$ has a periodicity along the magnetic axis defined as $\rho(z, \lambda) = \rho(z+2\pi/n_\mathrm{fp}, \lambda)$, where $n_\mathrm{fp}$ is the number of field periods. Therefore, an $\ell=2$ shaping mode with $k/n_\mathrm{fp} = \pm 1$ can be readily interpreted as an elongated shape that undergoes a $\pi$ rotation about the magnetic axis each field period. If $k/n_\mathrm{fp} = \pm 2$, the elongated shape rotates $2\pi$ per field period. The sign of $k$ then describes the direction of that rotation. Similar interpretations exist for all $\ell$ modes and any integer value of $k/n_\mathrm{fp}$.
	
	\section{Shape Analysis in Non-axisymmetric Equilibria \label{sec:non-axisymmetry}}
	
	In this section, three methods for characterizing the shape of the precise QA and QH configurations from Ref.~\cite{landreman_magnetic_2022} are presented. These methods differ in how they define a flux-surface cross-section, the coordinate $z$ along the magnetic axis, and the poloidal coordinate $\lambda$ used in Eq.~(\ref{eq:fourier_shape}). The utility of each method is determined by its ability to represent an equilibrium flux surface with a reduced number of non-negligible modes in the corresponding shape spectrum. 
	
	The first method uses planar cross-sections, has $z=\varphi_\mathrm{ma}$, and a translated geometric poloidal angle $\vartheta^{\prime}$ for $\lambda$. The translation of this angle is such that the $\vartheta^{\prime}=0$ and $\eta=0$ contours are equivalent, which ensures that the $\vartheta^{\prime}=0$ contour follows the minimal magnetic field strength on each flux surface. This is accomplished by defining two $\varphi_\mathrm{ma}$-dependent basis vectors
	\begin{align*}
		\hat{\mathbf{e}}^{\prime}_R(\varphi_\mathrm{ma}) &= \frac{\mathbf{r}_\mathrm{fs}(0, \xi^{\prime}) - \mathbf{r}_\mathrm{ma}(\varphi_\mathrm{ma})}{\| \mathbf{r}_\mathrm{fs}(0, \xi^{\prime}) - \mathbf{r}_\mathrm{ma}(\varphi_\mathrm{ma}) \|}, \\
		\hat{\mathbf{e}}^{\prime}_Z(\varphi_\mathrm{ma}) &= \hat{\mathbf{e}}^{\prime}_R \times \hat{\mathbf{t}}_\mathrm{ma}(\varphi_\mathrm{ma}),
	\end{align*}
	where $\xi^{\prime}$ is the value of $\xi$ at which the $\eta=0$ contour intersects the planar cross-section centered about $\varphi_\mathrm{ma}$, and $\mathbf{t}_\mathrm{ma} = \partial \mathbf{r}_\mathrm{ma} / \partial \varphi_\mathrm{ma}$ is the vector tangent to the magnetic axis, with the hat denoting a unit-vector normalization. The translated poloidal angle is then defined as 
	\begin{equation}
		\vartheta^{\prime} = \arctan\left( \frac{\boldsymbol{\rho} \cdot \hat{\mathbf{e}}^{\prime}_Z}{\boldsymbol{\rho} \cdot \hat{\mathbf{e}}^{\prime}_R } \right).
	\end{equation}
	Note that this method describes the conventional method for describing an axisymmetric equilibrium, generalized to a non-axisymmetric equilibrium. It is shown in Secs.~\ref{sec:QA_shaping} and \ref{sec:QH_shaping} that this method results in the largest number of non-negligable shaping modes in the shape spectrum of all three methods considered.
	
	The second method also utilizes the planar cross-sections, but with $z=\phi_\mathrm{ma}$ and $\lambda=\eta$. It is found empirically that this allows for a more uniform distribution in the poloidal angle within each cross-section, resulting in a lower number of non-negligible shaping modes relative to the previous method. The third method uses the torsioned cross-sections with $z=\phi_\mathrm{ma}$ and $\lambda=\eta$. This is shown to result in the lowest number of non-negligable shaping modes in the shape spectrum of all three methods considered. 
	
	To compute the precise QH and QA equilibria, the Variational Moment Equilibrium Code (VMEC) \cite{hirshman_1983} is used. VMEC is run in fixed-boundary mode with the pressure and current profiles set to zero, resulting in vacuum-field configurations. To then compute the Boozer angle coordinates needed to define $\eta$ and $\xi$, the BOOZ\_XFORM code \cite{booz_transform} is used. In the precise QA configuration, 16 poloidal and 12 toroidal modes are used in VMEC, and 96 poloidal and 48 toroidal modes are used in the BOOZ\_XFORM transformation. In the precise QH configuration, 8 poloidal and 8 toroidal modes are used in VMEC, and 48 poloidal and 40 toroidal modes are used in the BOOZ\_XFORM transformation. The $\eta$ and $\xi$ coordinates are then constructed using Eqs.~(\ref{eq:eta_def}) and (\ref{eq:xi_def}), where $(M, N)=(1, 0)$ in the QA configuration and $(1, -4)$ in the QH configuration. For additional details regarding how the $\eta$ and $\xi$ coordinates and their corresponding basis vectors are computed, see Appendix \ref{sec:sym_diffs}. For details regarding how $\phi$ is approximated on the magnetic axis, see Appendix \ref{sec:p_axis}.
	
	\subsection{Precise QA \label{sec:QA_shaping}}
	
	\begin{figure}
		\centering
		\includegraphics[width=.48\textwidth, keepaspectratio]{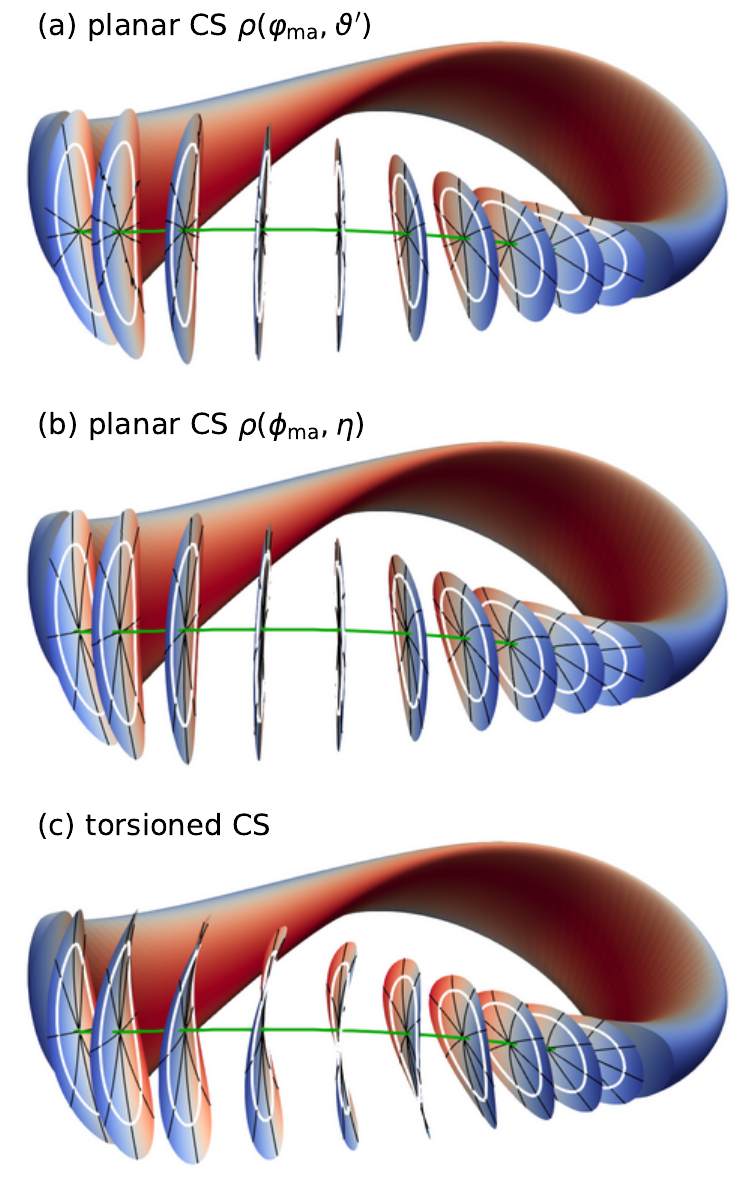}
		\caption{The 2D cross-sections are shown in the precise QA configuration for the planar cross-sections with $\rho(\varphi_\mathrm{ma}, \vartheta^{\prime})$ and $\rho(\phi_\mathrm{ma}, \eta)$ in panels (a) and (b), respectively, and the torsioned cross-sections in panel (c). The black curves follow the poloidal-coordinate contours and the white curve follows the $\psi/\psi_\mathrm{edge}=0.5$ flux surface. This shows that the torsioned cross-sections efficiently fill out the equilibrium volume by defining a uniformly spaced coordinate system. \label{fig:QA_CS}}
	\end{figure}
	
	The first configuration considered is the precise QA. In Fig.~\ref{fig:QA_CS}, the planar cross-sections are shown in panels (a) and (b) and the torsioned cross-sections are shown in panel (c). Along each cross-section, the black curves represent contours of the cross-section's poloidal angle $\lambda$ and the white curves denote the $\psi/\psi_\mathrm{edge}=0.5$ flux surface. In panel (a), where $\rho = \rho(\varphi_\mathrm{ma}, \vartheta^{\prime})$, one may observe the intuitive fact that using a geometric poloidal angle results in a clustering of the $\vartheta^{\prime}$ contours in regions where the flux-surface cross-section radius $\rho$ is small. By comparison, in panel (b), where $\rho=\rho(\phi_\mathrm{ma}, \eta)$, this clustering is greatly reduced, meaning that a more uniform distribution in the poloidal coordinate is achieved along each flux-surface cross-section. This was anticipated from Fig.~\ref{fig:sym_coords}, where the real-space clustering in the $\eta$ contours was observed in regions where the flux-surface gradient of $B$ was large. Here, it can be seen that this clustering results in a more uniform distribution in $\eta$ along a flux-surface cross-section relative $\vartheta^{\prime}$.
	
	In some of the panel (b) cross-sections, however, the radial $\eta$ contours can be observed to follow a curved path along the 2D cross-sections. This is especially true for the cross-sections closer to the right-hand side of the panel. This is in contrast to the torsioned cross-sections in panel (c), where, by construction, the radial $\eta$ contours follow the shortest path between their origin on the magnetic axis and the plasma edge. This is observed as a reduction in the curvature of the radial $\eta$ contours along each cross-section. Notably, any remaining curvature in the $\eta$ contours can be attributed to either a curvature of the $\eta$ manifold to which they adhere, or a curvature of the contour on that manifold in service to its path minimization between the magnetic axis and the plasma edge. This demonstrates how the torsioned cross-sections efficiently fill out the radial volume in a quasi-symmetirc equilibrium while adhering to uniformly distributed lines of symmetry.
	
	\begin{figure*}
		\centering
		\includegraphics[width=0.98\textwidth, keepaspectratio]{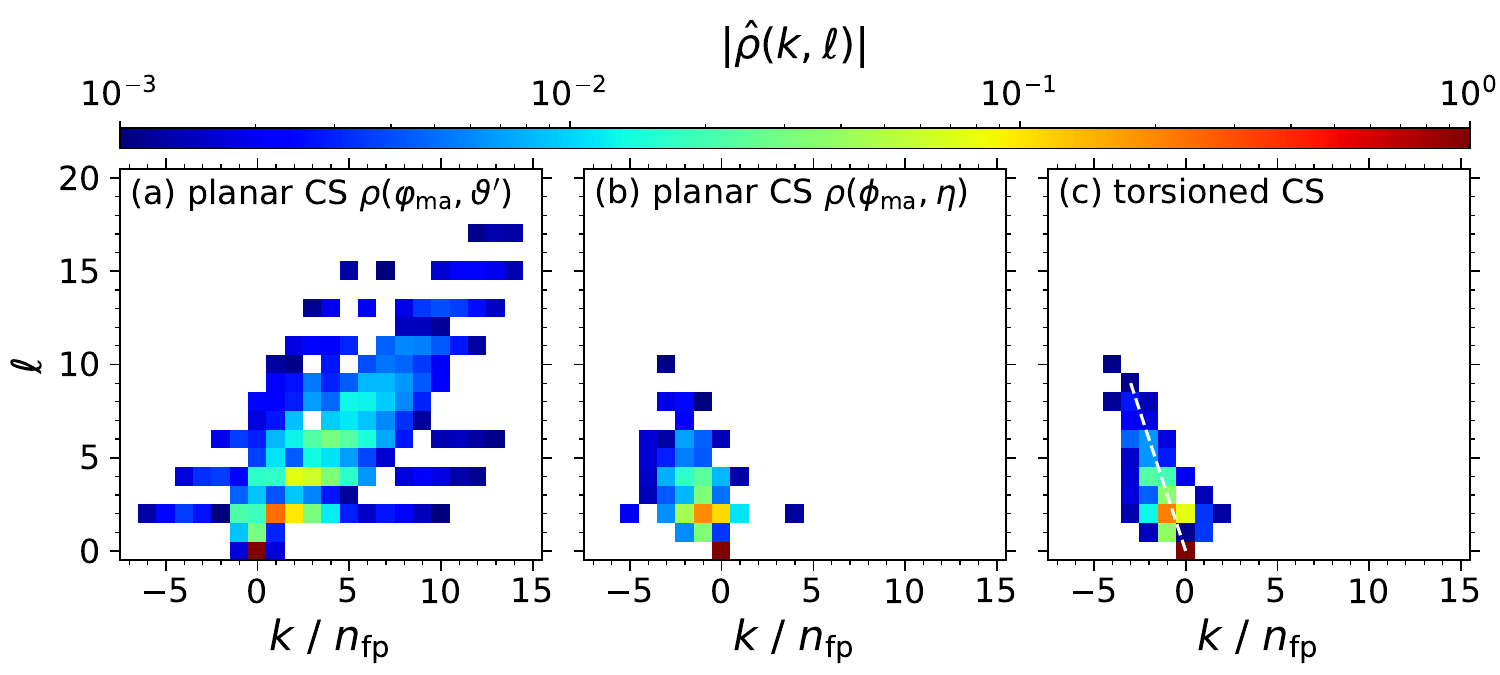}
		\caption{Shape spectra of the precise QA configuration are shown for the planar cross-sections with $\rho(\varphi_\mathrm{ma}, \vartheta^{\prime})$ and $\rho(\phi_\mathrm{ma}, \eta)$ in panels (a) and (b), respectively, and the torsioned cross-sections in column (c). This shows that the torsioned cross-sections reduce the amount of spectral content relative to the planar cross-sections and reveals a linear distribution of shape amplitudes, as indicated by the dashed white line in panel (c). \label{fig:QA_spec_comp}}
	\end{figure*}
	
	For each set of cross-sections, the $\psi/\psi_\mathrm{edge}=0.5$ flux surface is selected to show the shaping modes that result from the corresponding shape analysis. Repeating this analysis at different values of $\psi/\psi_\mathrm{edge}$ does change the spectra quantitatively, but the qualitative trends are the same. This is also true for the corresponding analysis on the precise QH configuration in Sec.~\ref{sec:QH_shaping}. The QA data are shown in Fig.~\ref{fig:QA_spec_comp}, where $|\hat{\rho}|$, normalized to $|\hat{\rho}(0, 0)|$, is shown as a function of $\ell$ and $k$. Panel (a) shows the data for the planar cross-sections with $\rho(\varphi_\mathrm{ma}, \vartheta^{\prime})$, panel (b) shows the data for the planar cross-sections with $\rho(\phi_\mathrm{ma}, \eta)$, and panel (c) shows the data for the torsioned cross-sections. 
	
	The color shows $|\hat{\rho}|$ on a logarithmic scale that spans three orders of magnitude, from $10^{-3}$ to $1$. It is found that reducing the lower bound on the color scale to values below $10^{-3}$ results in the visual appearance low-amplitude modes that span all values of $k$ at large $\ell$, and broadens the spectrum in $k$ at low $\ell$. If, however, one increases the number of grid points on a flux surface when computing the $\eta$ and $\xi$ coordinates, then the $|\hat{\rho}|$ threshold at which these low amplitude modes appear is reduced and values of $|\hat{\rho}|$ above that threshold remain unchanged. This is a strong indication that these low amplitude modes can be attributed to discretization noise. In this and all other shape spectra considered, $501$ poloidal and $501$ toroidal grid points are used in both VMEC and BOOZ\_XFORM to generate the $\eta$ and $\xi$ coordinates. This limits the relevant analysis to normalized shaping modes for which $|\hat{\rho}| \geq 10^{-3}$, with modes below this threshold assumed to be negligible.
	
	From Fig.~\ref{fig:QA_spec_comp}, it can be observed that the largest number of non-negligible shaping modes occurs in the $\rho(\varphi_\mathrm{ma}, \vartheta^{\prime})$ planar cross-sections. This is exemplified in panel (a) by the large number of shaping modes compared to panels (b) and (c). This is consistent with the interpretation of Fig.~\ref{fig:QA_cross_section_demo}, where it is argued that the clustering of the geometric poloidal angle in regions of small $\rho$ results in an inefficient representation of an equilibrium shape.	Turning one's attention to panels (b) and (c), there is a further reduction in the number of shaping modes going from the planar cross-sections to the torsioned cross-sections, though this reduction is less pronounced than the reduction from panel (a). 
	
	A notable change, however, can be observed in the distribution of shaping modes between panels (b) and (c). Namely, in panel (c), the torsioned cross-sections can be seen to produce a spectrum that is more tightly constrained about a characteristic diagonal compared to the distribution seen in panel (b). This distribution is indicated by the white dashed line in panel (c), which passes through the points $(k/n_\mathrm{fp}, \ell) = (0, 0)$ and $(-3, 9)$, following a slope of $-3$. Notably, there is a linear distribution of shaping modes in panels (a) and (b), but these spectra are polluted with a greater number of intermediate to low amplitudes modes. This suggests the unexpected result that, in addition to efficiently filling out the radial domain of a quasi-symmetric equilibrium, these torsioned cross-sections can also reveal a considerable amount of spatial regularity on a single flux surface with minimal spectral noise. For the precise QA configuration considered here, the distribution of shaping modes in panel (c) reveals that shapes with increasing complexity, identified as an increase in $\ell$, rotate about the magnetic axis with increasing speed, identified by an increase in $|k|/n_\mathrm{fp}$, and that the rate of this increase occurs at a ratio of approximately 3 to 1.
	
	\subsection{Precise QH \label{sec:QH_shaping}}
	
	\begin{figure}
		\centering
		\includegraphics[width=.48\textwidth, keepaspectratio]{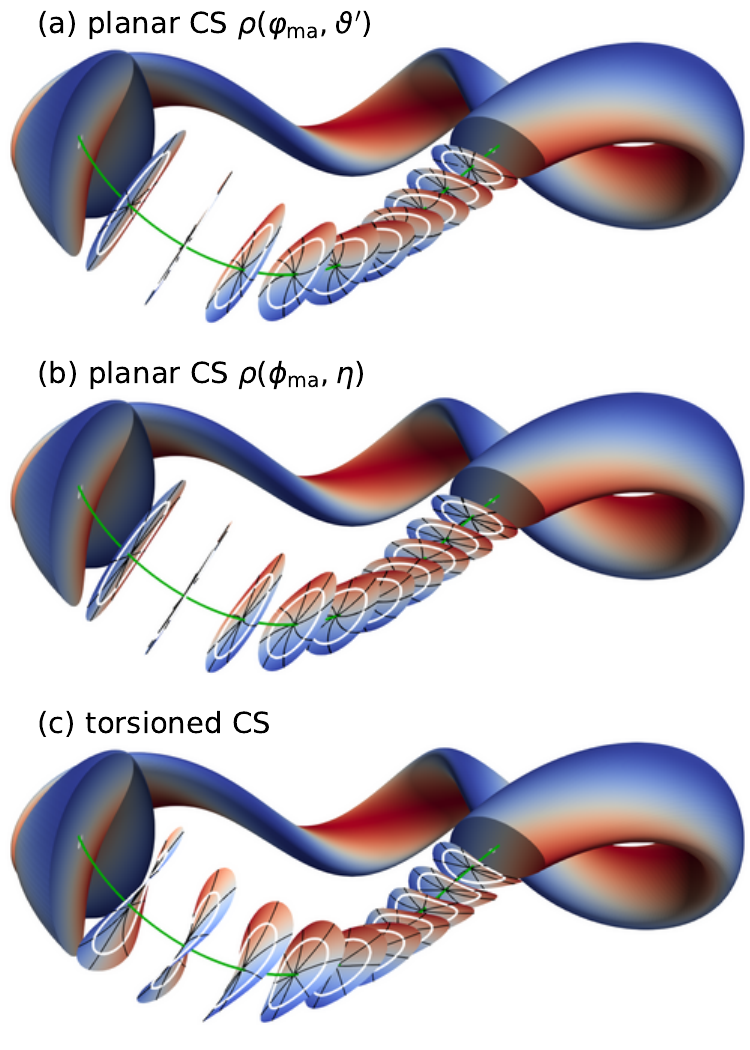}
		\caption{The 2D cross-sections are shown in the precise QH configuration for the planar cross-sections with $\rho(\varphi_\mathrm{ma}, \vartheta^{\prime})$ and $\rho(\phi_\mathrm{ma}, \eta)$ in panels (a) and (b), respectively, and the torsioned cross-sections in panel (c). The black curves follow the poloidal-coordinate contours and the white curve follows the $\psi/\psi_\mathrm{edge}=0.5$ flux surface. This shows that the torsioned cross-sections efficiently fill out the equilibrium volume by defining a uniformly spaced coordinate system. \label{fig:QH_CS}}
	\end{figure}
	
	Figure \ref{fig:QH_CS} shows the planar cross-sections of the precise QH configuration in panels (a) and (b) and the torsioned cross-sections in panel (c). As with the QA configuration in Fig.~\ref{fig:QA_CS}, the black radial curves along each cross-section show the radial poloidal angle contours and the white curve follows the $\psi/\psi_\mathrm{edge}=0.5$ flux surface. Consistent with Fig.~\ref{fig:QA_cross_section_demo}, panel (a) shows that when $\lambda=\vartheta^{\prime}$, the poloidal angle contours cluster in regions with relatively small $\rho$. By comparison, panel (b), where $\lambda=\eta$, shows a more uniform distribution in poloidal angle contours along the planar cross-sections. Then, transitioning to the torsioned cross-sections in panel (c), one can observe that the curvature in the radial $\eta$ contours is considerably reduced relative to those in panel (b), which is attributed to the path minimization that results from how the torsioned cross-sections are defined.
	
	\begin{figure*}
		\centering
		\includegraphics[width=0.98\textwidth, keepaspectratio]{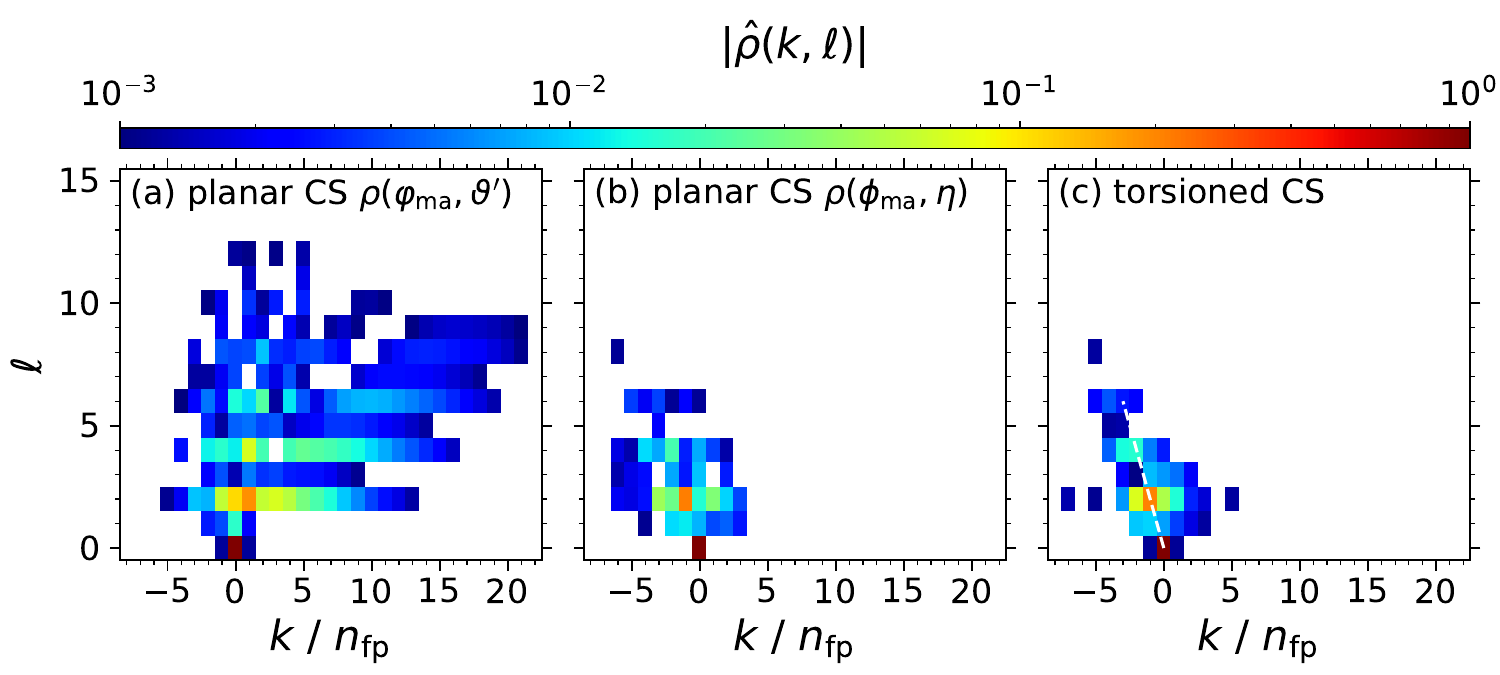}
		\caption{Shape spectra of the precise QH configuration are shown for the planar cross-sections with $\rho(\varphi_\mathrm{ma}, \vartheta^{\prime})$ and $\rho(\phi_\mathrm{ma}, \eta)$ in panels (a) and (b), respectively, and the torsioned cross-sections in column (c). This shows that the torsioned cross-sections reduce the amount of spectral content relative to the planar cross-sections and reveals a linear distribution of shape amplitudes, as indicated by the dashed black line in panel (c). \label{fig:QH_spec_comp}}
	\end{figure*}
	
	Figure \ref{fig:QH_spec_comp} shows the corresponding shape spectrum for all three cross-section sets, with the $\rho(\varphi_\mathrm{ma}, \vartheta^{\prime})$ planar cross-sections shown in panel (a), the $\rho(\phi_\mathrm{ma}, \eta)$ planar cross-sections in panel (b), and the torsioned cross-sections in panel (c). As with the QA configuration, it can be observed that the $\rho(\varphi_\mathrm{ma}, \vartheta^{\prime})$ planar cross-sections exhibit the largest number of non-negligable shaping modes of all three cross-section sets. That number is significantly reduced simply by transforming $\varphi_\mathrm{ma} \rightarrow \phi_\mathrm{ma}$ and $\vartheta^{\prime}\rightarrow\eta$, as demonstrated in column (b). 
	
	As in Fig.~\ref{fig:QA_spec_comp}, the torsioned cross-sections in panel (c) reveal a distribution of shaping modes about a characteristic diagonal. For the precise QH equilibrium, however, this line is shown to pass through the points $(k/n_\mathrm{fp}, \ell)=(0, 0)$ and $(-3, 6)$, following a slope of $-2$. This indicates that a similar relationship between shape complexity and shape rotation about the magnetic axis, as observed in Fig.~\ref{fig:QA_spec_comp}(c), also exists in the QH configuration, but here that rate of increase occurs at a ratio of 2 to 1. While diagonal distributions can be observed in panels (a) and (b), these distributions are polluted with a greater amount of spectral content relative to panel (c). 
	
	Due to the high degree of quasi-symmetry observed in both the QA and QH configurations, it is conjectured that this linear distribution of mode amplitudes in an equilibrium's shape spectrum is a consequence of its quasi-symmetry. Moreover, it is argued that the torsioned cross-sections, with $\eta$ used as a poloidal angle, makes this spatial regularity more salient. This is most likely due to the ability of the torsioned cross-sections to efficiently fill out the volume of a quasi-symmetric equilibrium. Evidence for these claims is provided by examining equilibrium shapes throughout the QUASR database.
	
	\section{Shaping Across the QUASR Database \label{sec:QUASR}}
	
	The QUASR database provides over $3.7\times 10^5$ stellarator geometries, with approximately 54\% of them QA and the remaining QH \cite{giuliani_2024_13717741}. Therefore, this database provides a useful dataset with which the stellarator shaping trends identified in Sec.~\ref{sec:non-axisymmetry} can be investigated. A major shortcoming in the data set used here is that approximately 30\% of the database failed to converge to a solution in VMEC. The reason for this is likely due to the large peak torsion in the magnetic axis of the failed configurations. This is a problem for VMEC because the code uses a cylindrical coordinate system for its real-space map, which is not well suited for configurations in which the magnetic axis is significantly displaced from the azimuthal plane. Many of the configurations that failed to converge are QH equilibria with $n_\mathrm{fp} \geq 6$. Other ideal-MHD equilibrium codes like DESC \cite{Dudt_2020} or GVEC \cite{gvec_2025_zenodo} may have a higher success rate when generating these highly torsioned equilibria.
	
	Restricting the present analysis to the converged set of equilibria, this section reports on two features in the distribution of shape amplitudes $|\hat{\rho}|$ in all shape spectra from across the converged set of equilibria. Shape spectra are computed for the $\psi/\psi_\mathrm{edge}=0.5$ flux surface, with the $\eta$ and $\xi$ coordiantes defined using $(M, N) = (1, 0)$ in all QA configurations and $(1, -n_\mathrm{fp})$ in all QH configurations. Features are compared for all three cross-section methods and investigated for their sensitivity to the quality of quasi-symmetry. Quasi-symmetry quality is defined as the the root-sum-square of the symmetry breaking modes in the Boozer spectrum
	\begin{equation*}
		\mathcal{Q}_\mathrm{err} = \frac{1}{B_{0,0}} \left(\sum\limits_{\substack{n\neq iN \\ m\neq iM}} B^2_{n,m}\right)^{1/2}.
	\end{equation*}
	Here, $n$ and $m$ are the toroidal and poloidal mode numbers, $B_{n,m}$ the corresponding Boozer mode amplitudes, $B_{0,0}$ is the flux-surface-averaged magnetic field strength, and $i$ is any finite and positive integer.
	
	The first shape spectrum feature considered is how many mid- to high-amplitude shaping modes are found in each of the three shape spectra. Here, mid- to high-aplitude modes are defined to be $10^{-2} \leq |\hat{\rho}| \leq 1$, spanning two orders of magnitude while remaining a factor of ten above the noise floor identified in Sec.~\ref{sec:QA_shaping}. 
	
	The second feature is how linear the distribution of shape amplitudes is in a spectrum. This is measured using the eigenvalues of a covariance matrix, where the matrix is constructed from the weighted sum of the variance of the variables $\ell$ and $k$, with the weights defined by the shape amplitudes. To define this procedure explicitly, the shape amplitude weights are defined as
	\begin{equation*}
		w(\ell, k) = \frac{|\hat{\rho}(k, \ell)|}{\sum_{\ell, k} |\hat{\rho}(k, \ell)|},
	\end{equation*}
	where the $k/n_\mathrm{fp} \in \left[-30, 30\right]$ and $\ell \in [0, 30]$ modes are included in the summation. A weighted average is then defined as
	\begin{align*}
		\langle \cdots \rangle = \sum_{\ell, k} \left( \cdots \right) w(\ell, k).
	\end{align*}
	The weighted variances are
	\begin{align*}
		\sigma_{kk} &= \langle \left( k - \langle k \rangle \right)^2 \rangle, \\
		\sigma_{k\ell} &= \langle \left(k - \langle k \rangle \right) \left(\ell - \langle \ell \rangle \right) \rangle, \\
		\sigma_{\ell\ell} &= \langle \left(\ell - \langle \ell \rangle \right)^2 \rangle,
	\end{align*}
	and the covariance matrix is
	\begin{equation*}
		\bar{\boldsymbol{\sigma}} =
		\begin{pmatrix}
			\sigma_{kk} & \sigma_{k\ell} \\
			\sigma_{k\ell} & \sigma_{\ell\ell}
		\end{pmatrix}.
	\end{equation*}

	Importantly, the eigenvectors of $\bar{\boldsymbol{\sigma}}$ define the axes about which the shape amplitudes are distributed, and the ratio of the corresponding eigenvalues---which are positive semi-definite---describe the proclivity of the shape amplitudes to be distributed about one eigenvector or the other. Therefore, taking the ratio of the minimum eigenvalue $\gamma_0$ over the maximum eigenvalue $\gamma_1$ provides a measure of how linear a shape amplitude distribution is, with values of $\gamma_0/\gamma_1 = 0$ describing a linear distribution and values of $\gamma_0/\gamma_1=1$ describing an isotropic distribution. This analysis is performed across the converged QUASR database, and includes all shaping mode amplitudes above the noise floor (i.e., $|\hat{\rho}| \geq 10^{-3}$).
	
	\begin{figure*}
		\centering
		\includegraphics[width=0.98\textwidth, keepaspectratio]{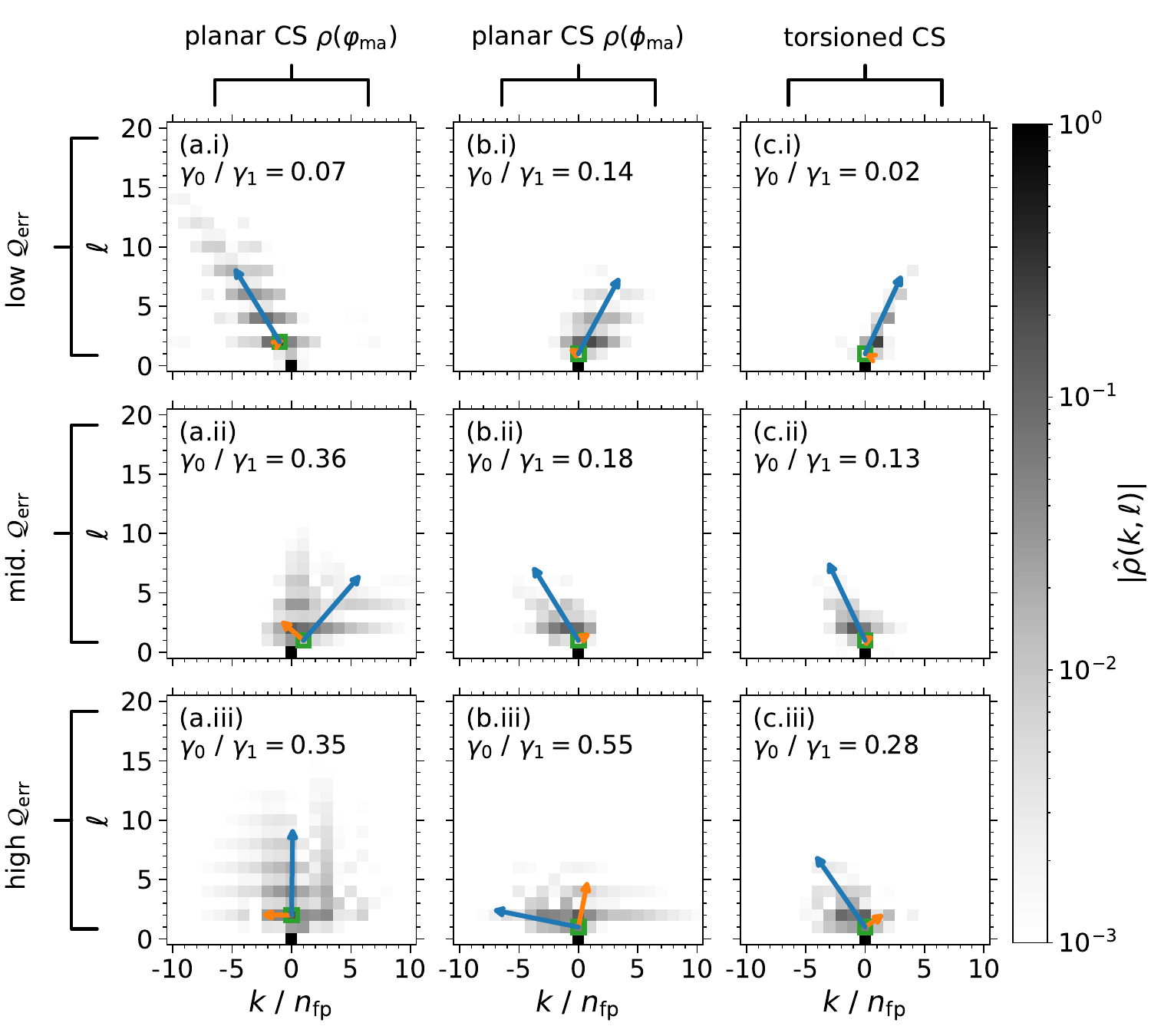}
		\caption{Shape spectra computed from three configurations of varying quasi-symmetry quality $\mathcal{Q}_\mathrm{err}$ are shown along each column. Across each row is shown the shape spectrum produced by each of the three methods for defining a cross-section. The arrows show the covariance matrix eigenvectors, with the blue arrow corresponding to the  $\gamma_1$ eigenvalue and the orange arrow corresponding to the $\gamma_0$ eigenvalue. The torsioned cross-sections reduce spectral content while also revealing the important relationship between shape rotation and shape complexity in producing quasi-symmetric equilibria.  \label{fig:covar_demo}}
	\end{figure*}
	
	To demonstrate these methods, Fig.~\ref{fig:covar_demo} shows the shape spectra for three different configurations using all three shaping methods. Each configuration's spectra are plotted across each row, for the low, mid., and high $\mathcal{Q}_\mathrm{err}$ configurations, with values of $\mathcal{Q}_\mathrm{err} = 5.3\times 10^{-4}$, $2.5\times 10^{-3}$, and $2.1\times 10^{-2}$, respectively. For reference, at $\psi/\psi_\mathrm{edge}=0.5$, in the precise QA and QH configurations, $\mathcal{Q}_\mathrm{err} = 7.8\times 10^{-2}$ and $2.7\times 10^{-4}$, respectively. The planar spectra with $\rho(\varphi_\mathrm{ma}, \vartheta^{\prime})$ are plotted along the left column, the planar spectra with $\rho(\phi_\mathrm{ma}, \eta)$ are plotted along the middle column, and the torsioned spectra are plotted along the right column. In each panel, the larger-eigenvalue eigenvector is represented with a blue arrow and the smaller-eigenvalue eigenvector with an orange arrow, with their relative magnitudes given by $\gamma_0/\gamma_1$. The shared origin between the eigenvectors is the $(\langle k\rangle, \langle \ell \rangle)$ point, outlined in green.
	
	Immediately, one can observe that the amount of spectral content is much greater in the left column than in the middle and right columns, with the right column showing the least amount of content across all rows. This is consistent with the observations from Sec.~\ref{sec:non-axisymmetry}, where the torsioned cross-sections were shown to reduce spectral content relative to the other two shaping methods. In the right column, one can observe that $\gamma_0/\gamma_1$ increases with $\mathcal{Q}_\mathrm{err}$, which suggests a correlation between a linear distribution of shape amplitudes and quasi-symmetry quality. These observations indicate that, in addition to reducing the amount of spectral content, the torsioned cross-sections also highlight an important relationship between shape rotation and shape complexity in producing a high degree of quasi-symmetry. In the remainder of this section, these trends are investigated throughout the QUASR database.

	\subsection{Quasi-axisymmetry in the QUASR database \label{sec:QUASR_QA}}
	
	The set of converged reconstructions from the QUASR database is comprised of 104,377 QA configurations that span $1 \leq n_\mathrm{fp} \leq 5$, $-0.964 \leq \stkout{\iota} \leq -8.55\times 10^{-4}$, and $2.30 \times 10^{-5} \leq \mathcal{Q}_\mathrm{err} \leq 0.544$. These configurations are split into three sets of equal size. The sets are ordered by quasi-symmetry quality, and are therefore labeled low, mid., and high $\mathcal{Q}_\mathrm{err}$, the low error set with $2.30 \times 10^{-5} \leq \mathcal{Q}_\mathrm{err} < 6.77\times 10^{-4}$, the mid.~error set with $6.77\times 10^{-4} \leq \mathcal{Q}_\mathrm{err} < 6.39\times 10^{-3}$, and the high error set with $6.39\times 10^{-3} \leq \mathcal{Q}_\mathrm{err} \leq 5.44$.
	
	\begin{figure*}
		\centering
		\includegraphics[width=0.98\textwidth, keepaspectratio]{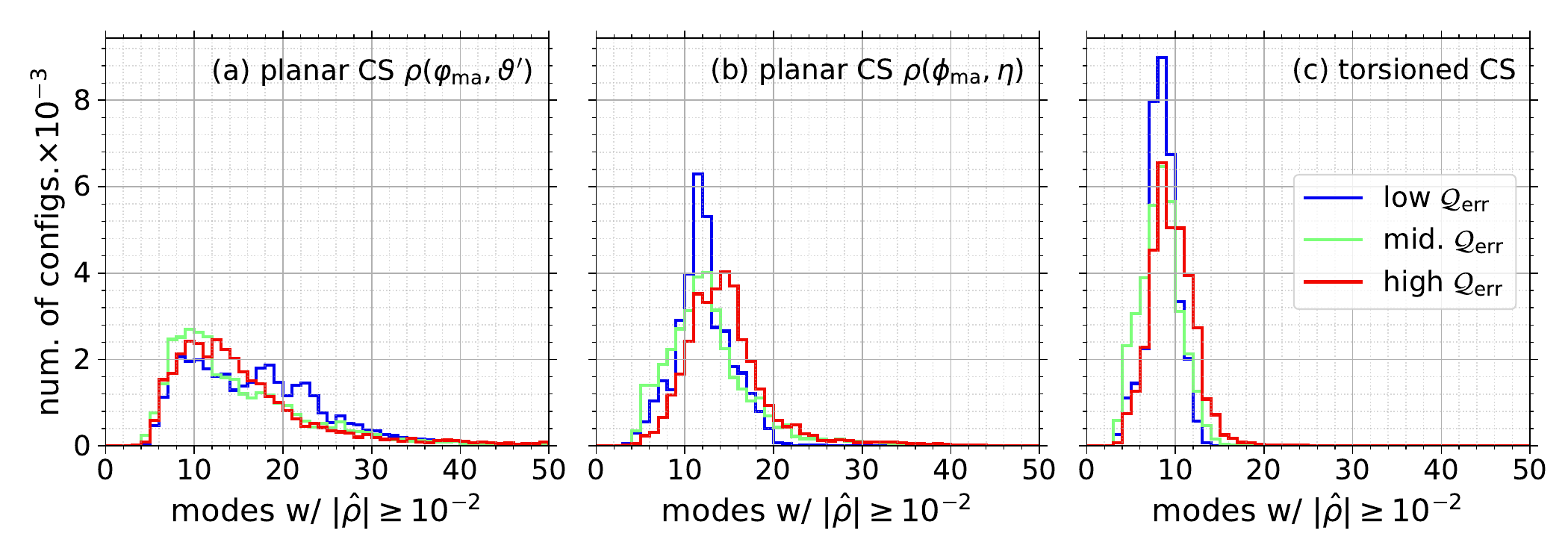}
		\caption{Histograms of the number of shaping modes with intermediate to high amplitudes for the planar with $\rho(\varphi_\mathrm{ma}, \vartheta^{\prime})$ (a), planar with $\rho(\phi_\mathrm{ma}, \eta)$ (b), and  torsioned (c) cross-sections. Sets of low, mid., and high $\mathcal{Q}_\mathrm{err}$ QA configurations are shown in blue, green, and red, respectively. This shows that the torsioned cross-sections significantly reduce the amount of spectral noise in a shape spectrum relative to the other shaping methods considered in QA equilibria, even when the quasi-symmetry is significantly degraded. \label{fig:QA_cells}}
	\end{figure*}

	Figure \ref{fig:QA_cells} shows histograms of the number of shape amplitudes in a spectrum that are above the $10^{-2}$ threshold. Histograms for the planar cross-sections with $\rho(\varphi_\mathrm{ma}, \vartheta^{\prime})$ are shown in panel (a), planar cross-sections with $\rho(\phi_\mathrm{ma}, \eta)$ are shown in panel (b), and torsioned cross-sections are shown in panel (c). In each panel, histograms for the low, mid., and high $\mathcal{Q}_\mathrm{err}$ sets are shown in blue, green, and red, respectively. Across all three panels, one can observe a clear trend in which the histogram peaks are increased, their widths are narrowed, and the distribution average is shifted to the left from panel (a) to (c). This is consistent with previous observations on the shape spectra for the precise QA configuration in Fig.~\ref{fig:QA_spec_comp}, where it was shown that the torsioned cross-sections produce the lowest amount of spectral content of all the shaping methods under consideration. Here, it is shown that this is a robust trend observed throughout the QUASR database of QA configurations.
	
	Interestingly, it is found here that this reduction persists even when the quasi-symmetry is significantly degraded. Panel (c), for example, shows that the average number of modes with amplitudes above $10^{-2}$ across all the high $\mathcal{Q}_\mathrm{err}$ configurations is approximately 9.5, with a standard deviation of 2.5. By comparison, the average number of modes across the same set of configurations for the planar cross-sections in panel (b) is approximately 14.5, with a standard deviation of 5. Continuing this trend, the average in panel (a) is approximately 15.5, with a standard deviation of 8. This shows that the torsioned cross-sections provide a robust method by which one can reduce the amount of spectral content in a shape spectrum of QA configurations, even when the quasi-symmetry quality is significantly degraded.
	
	If one further considers panel (c), it can be observed that the noise reduction is greatest in the low $\mathcal{Q}_\mathrm{err}$ configurations, for which the average number of modes is approximately 8.5 with a standard deviation of 1.5. These configurations exhibit higher peaks in their histogram than the mid.~and high $\mathcal{Q}_\mathrm{err}$ configurations, while the the mid.~and high $\mathcal{Q}_\mathrm{err}$ histograms can be observed to broaden relative to the low $\mathcal{Q}_\mathrm{err}$ set. Notably, the the mid.~$\mathcal{Q}_\mathrm{err}$ set is primarily broadened to include more configurations with a fewer number of $|\hat{\rho}| \geq 10^{-2}$ modes relative to the low $\mathcal{Q}_\mathrm{err}$ set. This is likely attributable to the fact that the quasi-symmetry error in the mid.~$\mathcal{Q}_\mathrm{err}$ configurations is significantly lower than that of the precise QA configuration presented in Sec.~\ref{sec:QA_shaping}. This means the mid.~$\mathcal{Q}_\mathrm{err}$ set includes many configurations that exhibit exceptional quasi-symmetry quality. Moreover, the sets of QA configurations considered here include equilibria with different distributions in their number of field periods and rotational transform. It is therefore speculated that these differences influence the amount of non-negligible spectral content within a given shape spectrum in configurations with a high-degree of quasi-symmetry, and only in configurations with a significant degradation in their quasi-symmetry does one observe a corresponding increase in spectral content, as can be observed in the high $\mathcal{Q}_\mathrm{err}$ histogram of panel (c). 
	
	\begin{figure*}
		\centering
		\includegraphics[width=0.98\textwidth, keepaspectratio]{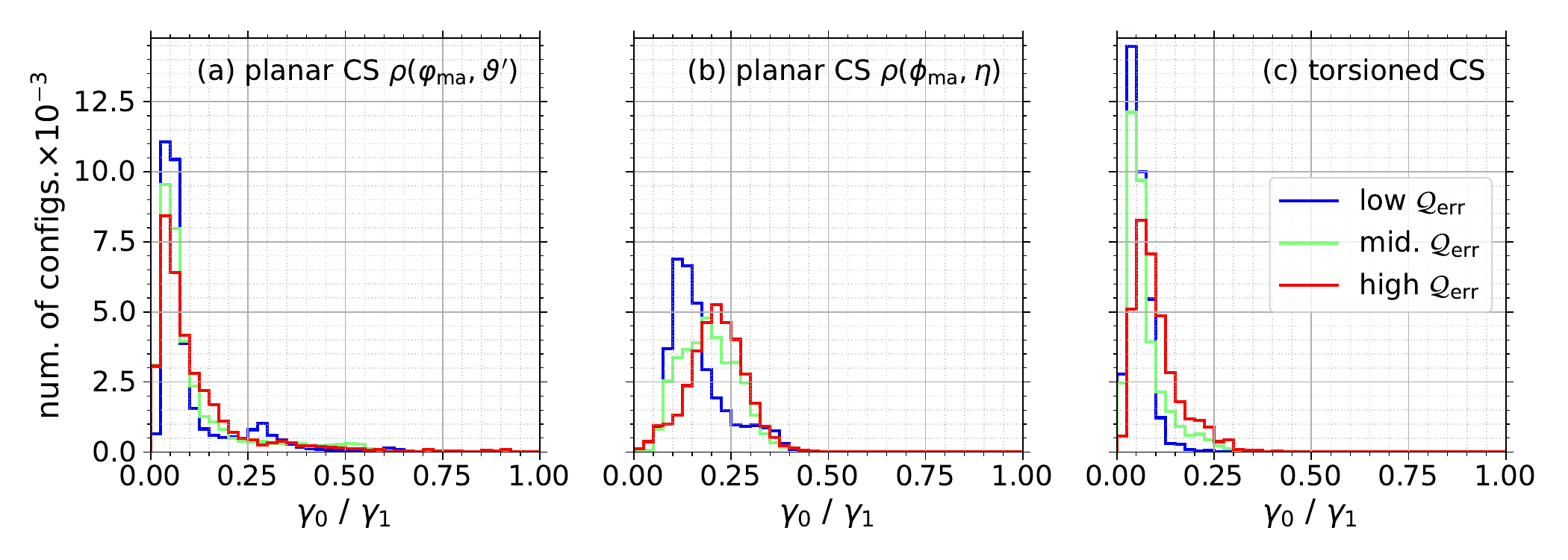}
		\caption{Histograms of the $\gamma_0/\gamma_1$ metric, which describes how linear a distribution of shape amplitudes is in a shape spectrum, in the planar with $\rho(\varphi_\mathrm{ma}, \vartheta^{\prime})$ (a), planar with $\rho(\phi_\mathrm{ma}, \eta)$ (b), and torsioned (c) cross-sections. Sets of low, mid., and high $\mathcal{Q}_\mathrm{err}$ QA configurations are shown in blue, green, and red, respectively. This suggests that a linear shape spectrum of the torsioned cross-sections is a necessary, but not sufficient condition for quasi-symmetry in QA configurations. \label{fig:QA_covar}}
	\end{figure*}

	Figure \ref{fig:QA_covar} shows histograms of the $\gamma_0/\gamma_1$ distributions from across the three shaping methods and the three $\mathcal{Q}_\mathrm{err}$ sets, with the layout of the figure identical to that of Fig.~\ref{fig:QA_cells}. Both the planar cross-sections in panel (a) and the torsioned cross-sections in panel (c) show a strong tendency towards a linear distribution of shape amplitudes, with their histogram peaks found at $\gamma_0/\gamma_1 < 0.1$. By comparison, the planar cross-sections in panel (b) are peaked below $\gamma_0/\gamma_1=0.25$ and exhibit a much broader distribution. All three panels, however, show a clear trend towards a more linearly distributed shape spectrum with increasing quasi-symmetry quality. This trend can be observed in panels (a) and (c) by the broadening of the histogram widths with degraded quasi-symmetry, and in panels (b) and (c) by a rightward shift in the histogram peaks with degraded quasi-symmetry. This implies that a linear distribution of shape spectrum amplitudes is a necessary, but not sufficient condition for good quasi-symmetry. Considered in conjunction with Fig.~\ref{fig:QA_cells}, one finds the torsioned cross-sections are able to reveal this dependence with a minimal number of spectral modes. 

	\subsection{Quasi-helical symmetry in the QUASR database \label{sec:QUASR_QH}}
	
	There are $112,745$ QH equilibria in the converged reconstructions from the QUASR database. These configurations span $2 \leq n_\mathrm{fp} \leq 8$, $-4.51\leq \stkout{\iota} \leq -0.495$, and $1.97\times 10^{-4} \leq \mathcal{Q}_\mathrm{err} \leq 1.09$. These configurations are split into three sets of equal size, as done previously for the QA configurations. Here, the low $\mathcal{Q}_\mathrm{err}$ set is with $1.97\times 10^{-4} \leq \mathcal{Q}_\mathrm{err} < 1.61\times 10^{-3}$, the mid.~$\mathcal{Q}_\mathrm{err}$ set is with $1.61 \times 10^{-3} \leq \mathcal{Q}_\mathrm{err} < 6.37\times 10^{-3}$, and the high $\mathcal{Q}_\mathrm{err}$ set is with $6.37\times10^{-3} \leq \mathcal{Q}_\mathrm{err} \leq 1.09$.
	
	\begin{figure*}
		\centering
		\includegraphics[width=0.98\textwidth, keepaspectratio]{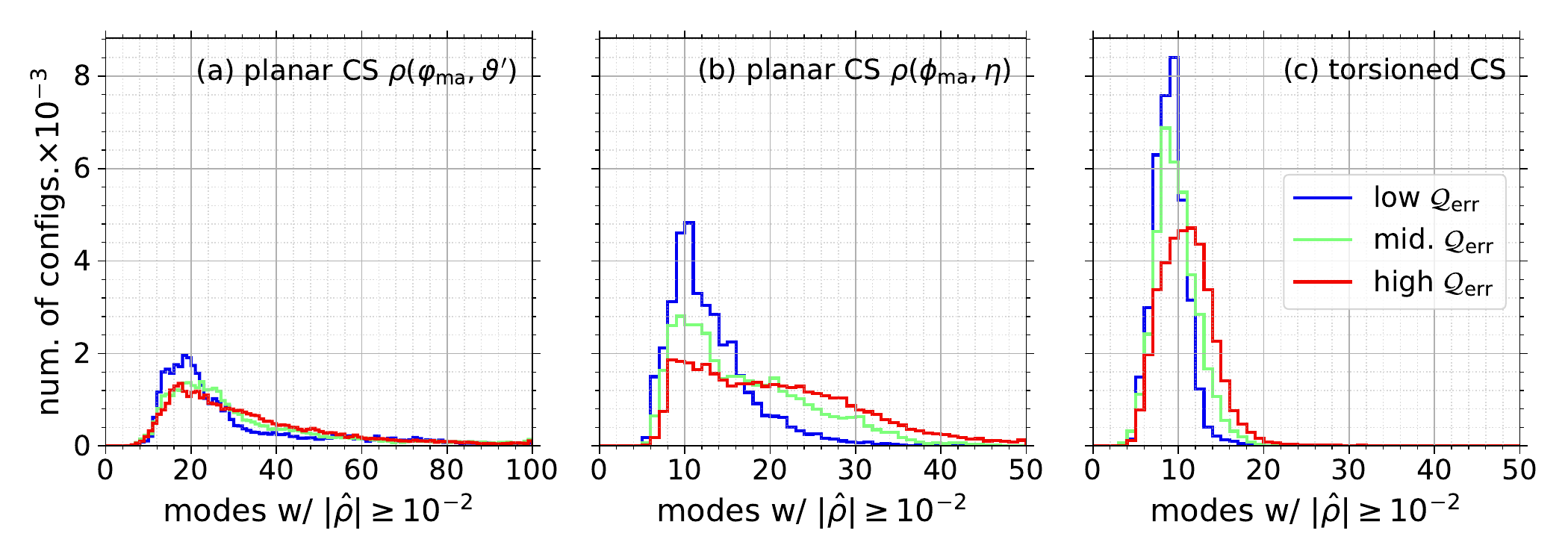}
		\caption{Histograms of the number of shaping modes with intermediate to high amplitudes for the planar with $\rho(\varphi_\mathrm{ma}, \vartheta^{\prime})$ (a), planar with $\rho(\phi_\mathrm{ma}, \eta)$ (b), and torsioned (c) cross-sections. Sets of low, mid., and high $\mathcal{Q}_\mathrm{err}$ QH configurations are shown in blue, green, and red, respectively. This shows that the torsioned cross-sections significantly reduce the amount of spectral content in a shape spectrum relative to the other shaping methods considered in QH equilibria, even when the quasi-symmetry is significantly degraded. \label{fig:QH_cells}}
	\end{figure*}

	Figure \ref{fig:QH_cells} shows histograms of the number of intermediate to high amplitude shaping modes from across all converged QH configurations. Similar to Fig.~\ref{fig:QA_cells}, the planar cross-sections with $\rho(\varphi_\mathrm{ma}, \vartheta^{\prime})$ are shown in panel (a), the planar cross-sections with $\rho(\phi_\mathrm{ma}, \eta)$ in panel (b), and the torsioned cross-sections in panel (c), while the low, mid., and high $\mathcal{Q}_\mathrm{err}$ sets are shown in blue, green, and red, respectively. Immediately, the same trend seen in Fig.~\ref{fig:QA_cells} can be observed here, namely that the number of spectral modes with $|\hat{\rho}| \geq 10^{-2}$ decreases from panel (a) to (b) to (c), as demonstrated by the leftward shift in the histogram peaks and the narrowing in the histogram widths as one moves from left to right. Note that the domain in panel (a) is twice that of panels (b) and (c), going from 100 to 50, respectively. The bin widths, however, are the same across all three panels. This further highlights the scale of the reduction in spectral content observed in the torisoned cross-sections.
	
	In all three panels, one can observe a clear dependence on the quality of quasi-symmetry, with the high $\mathcal{Q}_\mathrm{err}$ configurations producing a histogram that is broadened to include a greater number of $|\hat{\rho}| \geq 10^{-2}$ shaping modes relative to the low $\mathcal{Q}_\mathrm{err}$ configurations. Therefore, one can use the high $\mathcal{Q}_\mathrm{err}$ distribution as an upper bound on the number of mid.~to high amplitude modes needed to represent an equilibrium geometry. It is then notable that the high $\mathcal{Q}_\mathrm{err}$ histogram shows an average of 33.5 modes with a standard deviation of 19 in panel (a), an average of 20.5 and standard deviation of 9 in panel (b), and an average of 11 with a standard deviation of 3 in panel (c). This demonstrates the ability for the torsioned cross-sections to efficiently represent a QH equilibrium, even when the quality of that quasi-symmetry is significantly degraded.  
	
	\begin{figure*}
		\centering
		\includegraphics[width=0.98\textwidth, keepaspectratio]{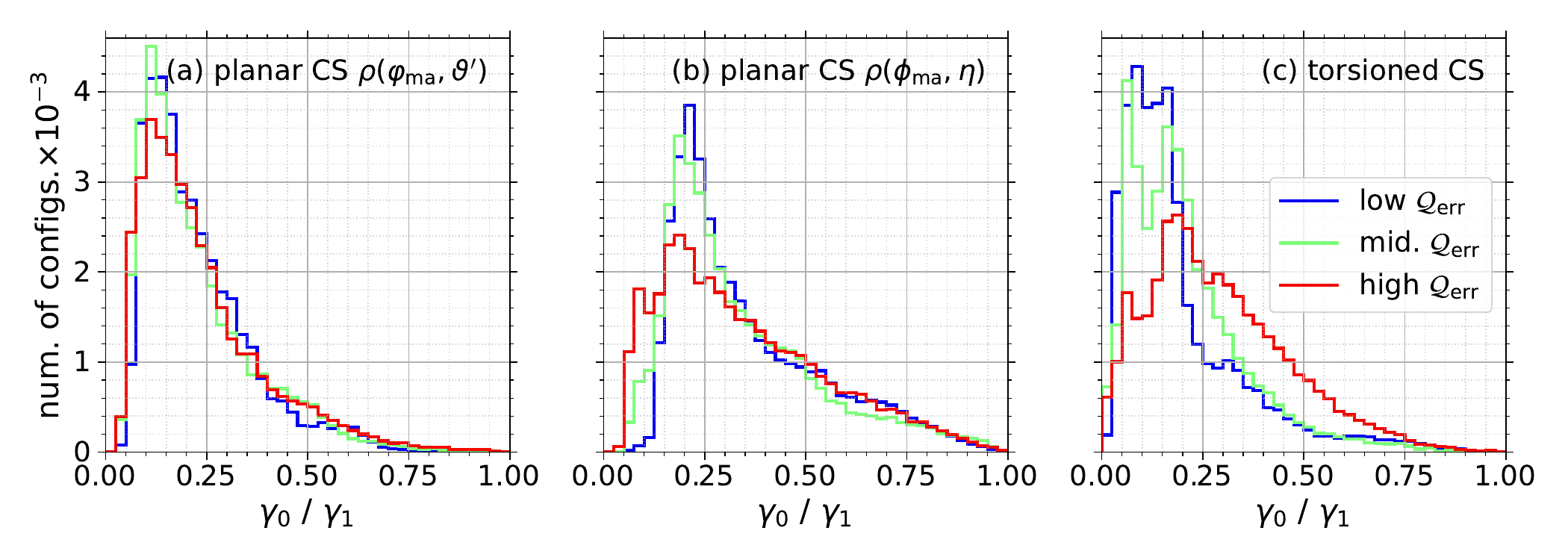}
		\caption{Histograms of the $\gamma_0/\gamma_1$ metric, which describes how linear a distribution of shape amplitudes is in a shape spectrum, in the planar with $\rho(\varphi_\mathrm{ma}, \vartheta^{\prime})$ (a), planar with $\rho(\phi_\mathrm{ma}, \eta)$ (b), and torsioned (c) cross-sections. Sets of low, mid., and high $\mathcal{Q}_\mathrm{err}$ QH configurations are shown in blue, green, and red, respectively. This shows that a linear distribution of shaping modes of the torsioned cross-sections is observed in a majority of the most quasi-symmetric QH configurations. \label{fig:QH_covar}}
	\end{figure*}

	Figure \ref{fig:QH_covar} shows the distribution of $\gamma_0/\gamma_1$ values across all converged QH equilibria, with the three shaping methods and quasi-symmetry sets presented in a layout identical to that of Fig.~\ref{fig:QA_covar}. The histogram peaks, across all three panels, occur at $\gamma_0/\gamma_1 < 0.25$, indicating a clear preference for more linearly distributed shape spectra. However, compared to Fig.~\ref{fig:QA_covar}, the histograms here have a broader distribution in $\gamma_0/\gamma_1$, indicating a stronger tendency towards more isotropic shape distributions relative to the QA configurations. Interestingly, compared to the QA configurations, the QH configurations do not show as strong of a dependence on quasi-symmetry quality in either planar cross-sections. This is not true, however, for the torsioned cross-sections in panel (c), where a broadening of the mid.~and high $\mathcal{Q}_\mathrm{err}$ histograms to include more isotropic shape distributions can be observed relative to the low $\mathrm{Q}_\mathrm{err}$ histogram. Moreover, the low $\mathcal{Q}_\mathrm{err}$ set of torsioned cross-sections produce a smaller number of $\gamma_0/\gamma_1 > 0.25$ configurations than is observed in either planar cross-sections. 
	
	Based on Fig.~\ref{fig:QA_covar}, it was argued that a linear distribution of shaping modes is a necessary but not sufficient condition for a high degree of quasi-symmetry in a QA equilibrium. Based on Fig.~\ref{fig:QH_covar}, an equivalent statement cannot be made regarding a QH equilibrium. However, Fig.~\ref{fig:QH_covar} does show that the torsioned cross-sections reveal a strong preference for a linearly distributed shape spectra when pursuing a high degree of quasi-symmetry in a QH equilibrium. 
	
	%It can be observed that the torsioned cross-sections exhibit the most sharply peaked $\gamma_0/\gamma_1$ distributions of all three shaping methods, with the peaks for all three $\mathcal{Q}_\mathrm{err}$ sets occurring just below $\gamma_0/\gamma_1=0.25$. Alternatively, the planar cross-sections both exhibit more isotropic spectral distributions, with many configurations producing values of $\gamma_0/\gamma_1 > 0.25$. Moreover, the torsioned cross-sections show the clearest dependence on quasi-symmetry quality. Similar to Fig.~\ref{fig:QA_covar}, the $\gamma_0/\gamma_1$ distributions shown here can be seen to broaden with increasing $\mathcal{Q}_\mathrm{err}$. Again, this indicates that a linear distribution in shaping mode amplitudes of the torisoned cross-sections is a necessary, but not sufficient condition for quasi-symmetry, exemplified here across a wide variety of QH equilibria.  
	
	\section{Conclusion \label{sec:conclusion}}
	
	Two methods for defining cross-sections in a quasi-symmetric ideal-MHD equilibrium have been presented. The first method is to define the cross-sections, at any point along the magnetic axis, as the intersection between an equilibrium flux-surface and a plane perpendicular to the magnetic axis at that point. These cross-sections are referred to as the planar cross-sections. The second method is to define a cross-section as the set of points across all quasi-symmetry contours that make their closest approach to the point on the magnetic axis about which the cross-section is centered. These cross-sections are referred to as the torsioned cross-sections. Importantly, both methods, when applied to an axisymmetric equilibrium, are shown to result in a poloidal cross-section. 
	
	A method for characterizing the equilibrium cross-section shapes, similar to methods used in previous publications, is presented. This method uses a Fourier transform along the poloidal  cross-section coordinate to produce shaping modes with familiar geometric interpretations (e.g., elongation, triangularity, squareness, etc.). Then, a Fourier transform along the magnetic axis results in shaping mode amplitudes that describe the rate at which the geometric cross-section shapes rotate about the magnetic axis. This 2D Fourier transform is referred to as a shape spectrum, and shape spectra are produced from both the torsioned and planar cross-sections, though the planar cross-sections are investigated using two different sets of Fourier-transformed coordinates.
	
	Analyzing the precise QA and QH configurations from Ref.~\cite{landreman_magnetic_2022}, it is found that the torsioned cross-sections produce shape spectra with significantly reduced spectral content relative to either spectra produced using the planar cross-sections. It is argued that this results from the ability of the torsioned cross-sections to efficiently fill out an equilibrium volume. It is also observed that the torsioned cross-sections produce shape spectra with a linear distribution of shape amplitudes in both the precise QA and QH configurations. 
	
	These trends are further investigated in the QUASR database of QA and QH configurations. It is shown empirically that, relative to the planar cross-sections, the torsioned cross-sections significantly reduce the spectral content in the shape spectra throughout the database. Using the torsioned cross-sections, configurations produce, on average, spectra with $\sim 10$ shaping modes of mid.~to high amplitude. This remains true even in configurations for which the quasi-symmetry quality is significantly degraded. This provides strong evidence that the torsioned cross-sections efficiently fill out an equilibrium volume, even when that equilibrium exhibits poor quasi-symmetry. 
	
	In addition to reducing the amount of spectral content, it is further shown that the shape spectrum of the torsioned cross-sections reveal a tight constraint on the distribution of non-negligible shape amplitudes in highly quasi-symmetric equilibria. It is found that high-quality QA configurations routinely exhibit a linear distribution of shape amplitudes, while high-quality QH configurations exhibit a strong preference for a linear distribution of shape amplitudes. This suggests a relationship between quasi-symmetry and equilibrium shaping, namely that increasingly complex shapes tend to undergo a proportional increase in their rotation about the magnetic axis when a high-degree of quasi-symmetry is observed.
	
	While the emperical nature of these observations prevent any conclusive statements, it is worth noting that the trends observed are based on observations from over $2\times 10^{5}$ QA and QH equilibria that span a wide range in their number of field periods, rotational transform, and quasi-symmetry quality. It is therefore argued that further effort should be made to clarify the relationship between the torsioned cross-sections, their corresponding shape spectra, and the quality of an equilibrium's quasi-symmetry. Moreover, future studies will focus on how the torsioned cross-sections might be used constructively in stellarator design, so that equilibria can be produced to achieve systematically modified shape spectra. The methods reported here provide an important first step in that direction. 
	
	\begin{acknowledgments}
		The authors thank E.~Miralles-Dolz, E.~Rodriguez, G.~Acton, and A.~Zettl for their insightful discussions. Support for this work was received through U.S. DOE Grant No.~DE-FG02-93ER54222.
	\end{acknowledgments}

	\appendix
	
	\section{Symmetry-aligned coordinates \label{sec:sym_diffs}}
	
	The VMEC coordinate system consists of the flux-surface label $s=\psi/\psi_\mathrm{edge}$, the poloidal angle $\theta_v$, and toroidal angle $\phi_v$, where the toroidal VMEC angle is equivalent to the negative azimuthal angle $\phi_v=-\varphi$. To transform the VMEC flux-surface coordinates into Boozer coordinates, one may use the relations
	\begin{align}
		\theta &= \theta_v + \lambda(s,\, \theta_v,\, \phi_v) - \stkout{\iota} p(s,\, \theta_v,\, \phi_v), \label{eq:pol_v2b} \\
		\phi &= \phi_v - p(s,\, \theta_v,\, \phi_v) \label{eq:tor_v2b}.
	\end{align}
	Here $\lambda$ is a periodic stream function that transforms VMEC coordinates to straight-field-line PEST coordinates and $p$ is a periodic stream function that transforms PEST coordinates to Boozer coordinates \cite{booz_transform}. 
	
	In VMEC, if one assumes stellarator symmetry, the stream function $\lambda$ is defined with the following Fourier series
	\begin{align*}
		\lambda &= \sum_{m=0}^\mathcal{M} \sum_{n=-\mathcal{N}}^\mathcal{N} L_{mn}(s) \sin(m\theta_v - nn_\mathrm{fp}\phi_v).
	\end{align*}
	Here, $L_{mn}$ is the radially-dependent Fourier amplitude and $\mathcal{M}$ and $\mathcal{N}$ are the number of poloidal and toroidal mode numbers included in the Fourier sum, respectively. Similarly, the function $p$ is defined in the BOOZ\_XFORM output as
	\begin{equation*}
		p = \sum_{m=0}^\mathcal{M_\mathrm{b}} \sum_{n=-\mathcal{N_\mathrm{b}}}^\mathcal{N_\mathrm{b}} P_{mn}(s) \sin(m\theta_v - nn_\mathrm{fp}\phi_v),
	\end{equation*}
	where the $\mathrm{b}$ subscripts denotes the corresponding BOOZ\_XFORM quantity. With these functions defined, it is straightforward to construct the symmetry-aligned coordinates $\eta$ and $\xi$ as defined in Eqs.~(\ref{eq:eta_def}) and (\ref{eq:xi_def}). 
	
	When computing the torsioned cross-sections, it is helpful to construct the covariant $\eta$ and $\xi$ basis vectors. A covariant basis vector is defined for the $i^\mathrm{th}$ coordinate as $\mathbf{e}_i = \partial\mathbf{r} / \partial \alpha_i$, where the $\alpha_j \neq \alpha_i$ coordinates are held fixed \cite{Dhaeseleer}. Since VMEC uses a cylindrical coordinate system, it is prudent to define $\mathbf{r}=R\hat{\mathbf{e}}_R + Z\hat{\mathbf{e}}_Z$, where $\hat{\mathbf{e}}_R$, $\hat{\mathbf{e}}_{\varphi}$, and $\hat{\mathbf{e}}_Z$ are the normalized cylindrical basis vectors. Then, the covariant VMEC basis vectors are defined as 
	\begin{align}
		\mathbf{e}_{s} &= \frac{\partial R}{\partial s} \hat{\mathbf{e}}_\mathrm{R} + \frac{\partial Z}{\partial s} \hat{\mathbf{e}}_\mathrm{Z}, \label{eq:covar_s} \\
		\mathbf{e}_{\theta} &= \frac{\partial R}{\partial \theta_v} \hat{\mathbf{e}}_\mathrm{R} + \frac{\partial Z}{\partial \theta_v} \hat{\mathbf{e}}_\mathrm{Z}, \label{eq:covar_u} \\
		\mathbf{e}_{\phi} &= \frac{\partial R}{\partial \phi_v} \hat{\mathbf{e}}_\mathrm{R} + R\hat{\mathbf{e}}_{\varphi} + \frac{\partial Z}{\partial \phi_v} \hat{\mathbf{e}}_\mathrm{Z}, \label{eq:covar_v}
	\end{align}
	where the coordinates $R$ and $Z$ are defined in VMEC, assuming stellarator symmetry, as
	\begin{align*}
		R &= \sum_{m=0}^\mathcal{M} \sum_{n=-\mathcal{N}}^\mathcal{N} R_{mn}(s) \cos(m\theta_v - nn_\mathrm{fp}\phi_v),  \\
		Z &= \sum_{m=0}^\mathcal{M} \sum_{n=-\mathcal{N}}^\mathcal{N} Z_{mn}(s) \sin(m\theta_v - nn_\mathrm{fp}\phi_v).
	\end{align*}
	
	To then compute the symmetry-aligned covariant basis vectors, one is required to compute the $\eta$ and $\xi$ derivatives. Expressions for these derivatives can be defined in terms of VMEC coordinates using the chain rule
	\begin{align*}
		\frac{\partial}{\partial \eta} &= \frac{\partial \theta_v}{\partial \eta} \frac{\partial}{\partial \theta_v} + \frac{\partial \phi_v}{\partial \eta} \frac{\partial}{\partial \phi_v}, \\ 
		\frac{\partial}{\partial \xi} &= \frac{\partial \theta_v}{\partial \xi} \frac{\partial}{\partial \theta_v} + \frac{\partial \phi_v}{\partial \xi} \frac{\partial}{\partial \phi_v}.
	\end{align*}
	Since derivatives taken with respect to some variable are performed while keeping the other variables within that coordinate system fixed, the following relations must be enforced 
	\begin{align*}
		\frac{\partial \eta}{\partial \eta} &= \frac{\partial \xi}{\partial \xi} = 1, &
		\frac{\partial \eta}{\partial \xi} &= \frac{\partial \xi}{\partial \eta} = 0.
	\end{align*}
	Therefore, taking the $\eta$ derivative of the $\eta$ and $\xi$ coordinates, with both expressed in VMEC coordinates, one may derive a system of equations that can be solved for $\partial \theta_v / \partial \eta$ and $\partial \phi_v / \partial \eta$. Similarly, taking the $\xi$ derivative of $\eta$ and $\xi$ will result in another system of equations that can be solved for $\partial \theta_v / \partial \xi$ and $\partial \phi_v / \partial \xi$. Note, however, a distinction must be made between a QA and QH equilibrium since $\xi$ has a different definition in each.
	
	Following the procedure thus described, the $\eta$ and  $\xi$ derivatives of the VMEC flux-surface coordinates are found to be
	\begin{align}
		\frac{\partial \theta_v}{\partial \eta} &= \frac{\mathcal{C}}{\Delta} \frac{\partial \xi}{\partial \phi_v}, \\
		\frac{\partial \phi_v}{\partial \eta} &= - \frac{\mathcal{C}}{\Delta} \frac{\partial \xi}{\partial \theta_v}, \\
		\frac{\partial \theta_v}{\partial \xi} &= -\frac{\mathcal{C}}{\Delta} \frac{\partial \eta}{\partial \phi_v}, \\
		\frac{\partial \phi_v}{\partial \xi} &= \frac{\mathcal{C}}{\Delta} \frac{\partial \eta}{\partial \theta_v}, \label{eq:dv_dxi}
	\end{align}
	with the Jacobian of the transformation from VMEC to Boozer coordinates defined as
	\begin{equation*}
		\Delta = \frac{\partial \theta}{\partial \theta_v} \frac{\partial \phi}{\partial \phi_v} - \frac{\partial \phi}{\partial \theta_v} \frac{\partial \theta}{\partial \phi_v},
	\end{equation*}
	and
	\begin{equation*}
		\mathcal{C} = 
		\begin{cases}
			1/M & \text{for QA} \\
			MN / \left(M^2 + N^2\right) & \text{for QH}
		\end{cases}.
	\end{equation*}
	Finally, using Eqs.~(\ref{eq:covar_u})--(\ref{eq:dv_dxi}), the symmetry-aligend covariant basis vectors are expressed as
	\begin{align}
		\mathbf{e}_{\eta} &= \frac{\mathcal{C}}{\Delta} \left( \frac{\partial \xi}{\partial \phi_v} \mathbf{e}_{\theta} - \frac{\partial \xi}{\partial \theta_v} \mathbf{e}_{\phi} \right), \label{eq:covariant_eta_QA} \\
		\mathbf{e}_{\xi} &= - \frac{\mathcal{C}}{\Delta} \left( \frac{\partial \eta}{\partial \phi_v} \mathbf{e}_{\theta} - \frac{\partial \eta}{\partial \theta_v} \mathbf{e}_{\phi} \right). \label{eq:covariant_xi_QA}
	\end{align}

	\section{Toroidal Boozer coordinate on the magnetic axis \label{sec:p_axis}}
	
	In BOOZ\_XFORM, the transfer function $p(s, \theta_v, \phi_v)$ is computed on the radial half-mesh. The half-mesh values of $s$ are defined as $s_i = (i + 0.5)/(n_s - 1)$, where $i$ is any integer from $[0, n_s-2]$ and $n_s$ is the number of flux surfaces on the VMEC full mesh. This means that $p$ is not computed on the magnetic axis and Eq.~(\ref{eq:tor_v2b}) cannot be used to compute $\phi_\mathrm{ma}$ directly. Instead, an extrapolated stream function $\tilde{p}(\phi_{v})$ is used, defined as
	\begin{equation*}
		\tilde{p}(\phi_{v}) = \frac{1}{L_{\theta}} \int\limits_{0}^{2\pi} p(s_0, \theta_v, \phi_{v}) \sqrt{g_{\theta\theta}} d\theta_v,
	\end{equation*}
	with
	\begin{equation*}
		L_{\theta}(\phi_v) = \int\limits_{0}^{2\pi} \sqrt{g_{\theta\theta}} d\theta_v,
	\end{equation*}
	and $g_{\theta\theta} = \mathbf{e}_{\theta} \cdot \mathbf{e}_{\theta}$. Therefore, $\tilde{p}(\phi_v)$ is a line average of $p$ along a poloidal cross-section at constant $\phi_v$ on the inner-most flux-surface of the radial half-mesh. To compute $\phi_\mathrm{ma}$, it is then assumed in Eq.~(\ref{eq:tor_v2b}) that $p(0, \theta_v, \phi_v) = \tilde{p}(\phi_v)$.
		
	\bibliographystyle{apsrev4-2}
	\bibliography{refs}
	
	\end{document}